\documentclass[preprint,12pt]{elsarticle}

\usepackage{amssymb}
\usepackage{amsmath}
\PassOptionsToPackage{hyphens}{url}\usepackage{hyperref}
\usepackage{subcaption}
\usepackage{multirow}
\usepackage{tabularx}
\newcolumntype{Y}{>{\centering\arraybackslash}X}
\usepackage{graphicx}

\journal{Minerals Engineering}

\begin{document}

\begin{frontmatter}



\title{AI-Driven Optimization under Uncertainty for Mineral Processing Operations} 

\author[mse,eps]{William Xu\corref{cor}}
\cortext[cor]{Corresponding author.}
\ead{williamx@stanford.edu}

\author[eps]{Amir Eskanlou}
\author[aa]{Mansur Arief}
\author[eps]{David Zhen Yin}
\author[eps]{Jef K.~Caers}

\affiliation[mse]{organization={Materials Science \& Engineering},
            addressline={Stanford University}, 
            city={Stanford},
            postcode={94305}, 
            state={CA},
            country={USA}}
\affiliation[eps]{organization={Earth \& Planetary Sciences},
            addressline={Stanford University}, 
            city={Stanford},
            postcode={94305}, 
            state={CA},
            country={USA}}
\affiliation[aa]{organization={Aeronautics \& Astronautics},
            addressline={Stanford University}, 
            city={Stanford},
            postcode={94305}, 
            state={CA},
            country={USA}}

\begin{abstract}
The global capacity for mineral processing must expand rapidly to meet the demand for critical minerals, which are essential for building the clean energy technologies necessary to mitigate climate change. However, the efficiency of mineral processing is severely limited by uncertainty, which arises from both the variability of feedstock and the complexity of process dynamics. To address this uncertainty, the current approach to designing and operating mineral processing circuits emphasizes process stability and control, relying on limited and/or indirect empirical tests, deterministic methods, and expert intuition. Yet a significant portion of valuable minerals is lost in waste streams, translating to millions of dollars of lost revenue and greater potential for environmental damage.

To optimize mineral processing circuits under uncertainty, we introduce an AI-driven approach that formulates mineral processing as a Partially Observable Markov Decision Process (POMDP). We demonstrate the capabilities of this approach in handling both feedstock uncertainty and process model uncertainty to optimize the operation of a simulated, simplified flotation cell as an example. We show that by integrating the process of information gathering (i.e., uncertainty reduction) and process optimization, this approach has the potential to consistently perform better than traditional approaches at maximizing an overall objective, such as net present value (NPV). We highlight the power of this approach in scenarios where the dynamics of the system, and subsequently the relationship between the inputs (e.g., feedstock composition, flotation operation settings) and desired outputs (e.g., recovery and grade), are not well known. Our methodological demonstration of this optimization-under-uncertainty approach for a synthetic case provides a mathematical and computational framework for later real-world application, with the potential to improve both the laboratory-scale design of experiments and industrial-scale operation of mineral processing circuits without any additional hardware.
\end{abstract}

\begin{graphicalabstract}
\includegraphics[width=\textwidth]{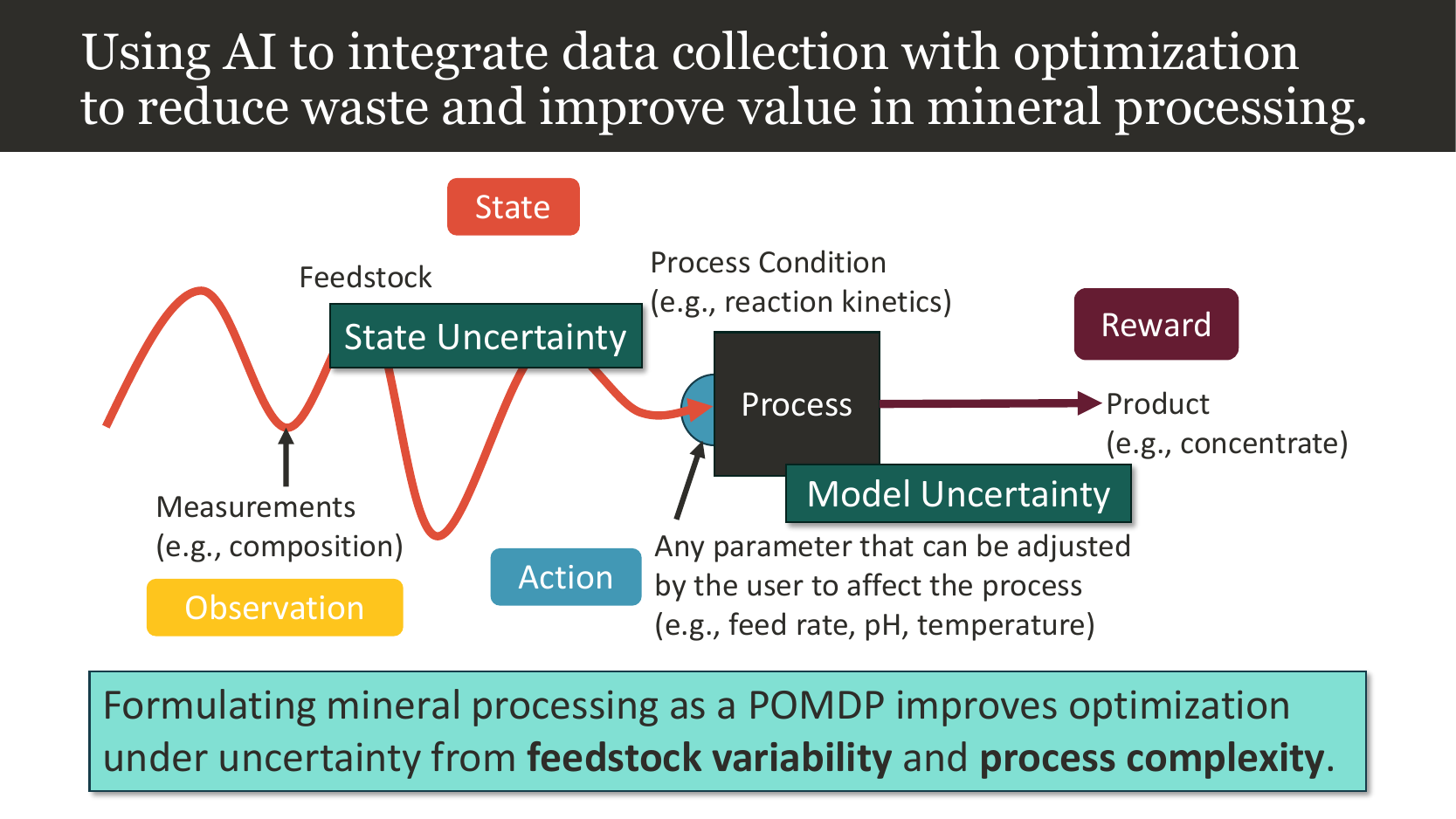}
\end{graphicalabstract}

\begin{highlights}
\item We frame mineral processing as a problem of optimization under uncertainty.
\item Uncertainty is represented as a belief of the mineral processing system.
\item An accurate model of the real system can be learned over time with belief updating.
\item We demonstrate the power of integrating uncertainty reduction with optimization.
\item This approach can be applied to optimize a wide range of mineral processing systems.

\end{highlights}


\end{frontmatter}



\section{Introduction}
\label{sec:intro}

In 2024, the average global temperature surpassed the 1.5°C threshold set by the UN Intergovernmental Panel on Climate Change (IPCC) for the first time in recorded history \cite{WMO2025}. The IPCC’s most recent report is clear: the consequences of human-caused climate change are immense and already taking place, and a clean energy transition is necessary to cut carbon emissions and mitigate these consequences \cite{lee2023synthesis}. To build the requisite clean energy technologies in such a short timeframe will require rapid sourcing of vast quantities of critical minerals \cite{iea2021criticalminerals}. At the same time, many countries have expressed geopolitical and national security concerns regarding critical mineral supply chains—especially for refining and processing capacity, which is heavily concentrated in China.

Mineral processing, a key component to the sourcing of critical minerals, faces increasing difficulties stemming from declining ore quality and a growing imperative to improve environmental performance. Additionally, mineral processing is energy and water intensive \cite{iea2021criticalminerals}, and so as an industry, it must work towards improving the sustainability of operations to ensure consistency with the eventual goal of combating climate change. Improving the efficiency of mineral processing can serve the dual goal of reducing waste and resource usage while increasing production and thereby revenue.

The efficiency of mineral processing is severely limited by uncertainty from both the variability of feedstock and the complexity of process dynamics \cite{amini2017optimization, amini2021design, koermeroptimization, koermer2022bayesian}. As \citet{bascur2019processcontrol} states, ``a critical problem in the process of ore extraction is the variability of the different elements that constitute the ore''. Traditionally, this uncertainty is addressed with process control and operational intelligence, which go hand-in-hand. Process control seeks to minimize variations in output by adjusting control parameters in response to input variations, and operational intelligence aims to collect information via real-time sensors to inform process control and optimization \cite{bascur2019processcontrol, bascur2024engineering}.

The proportional–integral–derivative (PID) control scheme is still the most commonly used process control technique today, but is only sufficient for single-loop systems, which have one controlled and one manipulated variable. Indeed, despite decades of research into advanced control, \citet{hodouin2011methods} and \citet{shean2011review} both emphasize that PID controllers continue to dominate industrial mineral processing, with limited measurable improvements in performance. Even where advanced multivariable or predictive control has been introduced, it is often constrained by the reliability of sensors, model uncertainty, and the difficulty of tuning nonlinear systems \cite{jovanovic2015contemporary}. These reviews collectively highlight that while model predictive control (MPC) has become the de facto ``advanced'' option, its effectiveness depends heavily on accurate, deterministic models and consistent process dynamics, which is an unrealistic representation of industrial flotation or grinding circuits. As a result, the traditional approach relies on a mix of expert intuition, empirical testing, extensive data collection, and deterministic optimization and control methods \cite{jiang2017data}.

Notably, \citet{hodouin2011methods} and \citet{shean2011review} each call for a more holistic or hierarchical view of process optimization, integrating sensors, observers, controllers, and optimizers. Yet, even in these ``optimization'' frameworks, optimization remains subordinate to control: setpoints are tuned to achieve a target grade or recovery, rather than directly optimizing the operation itself. In practice, this means the underlying objective functions---whether metallurgical or economic---are treated as supervisory layers above fixed control architectures, rather than as part of a unified decision process. As \citet{jovanovic2015contemporary} notes, the result is an architecture that can stabilize the process but struggles to adapt optimally when ore characteristics or process conditions shift unpredictably.

Accordingly, the most common approach to operating a mineral processing plant is to view it as a problem of control first and optimization second, even though the overarching goal of the plant is an optimization problem (i.e., maximizing economic profit, sustainability, safety, etc.). We will study the alternative: approaching process operation as a problem of optimization first and foremost, with subproblems of control. In this view, the goal is not to control variations, as process control seeks to do, but rather to optimize the process while accounting for variations---in other words, to leverage uncertainty rather than fight it. The potential value of optimization is undeniable \cite{bascur2019processcontrol, ding2012knowledge, hodouin2001state}, and the limitations of current control frameworks suggest that a probabilistic, decision-theoretic formulation may be necessary to achieve it.

A few projects have demonstrated the value of considering uncertainty in the optimization of mineral processing. \citet{valikangas2025evaluation} used sensitivity analysis and uncertainty propagation to understand the influence of feedstock variability and inform data collection. \citet{koch2020sequential} and Amini \cite{amini2017optimization, amini2021design} presented stochastic approaches that outperformed deterministic methods at designing mineral processing circuits. \citet{jiang2017data} and Koermer \citet{koermeroptimization, koermer2022bayesian} used reinforcement learning (RL) and machine learning (ML) to determine optimal operating conditions given unknown process dynamics at steady-state. This body of prior work forms a strong basis for considering uncertainty in optimizing mineral processing.

It is important to note that this paper focuses on decision-making and optimization under uncertainty, which is what RL is designed to do, rather than data-driven modeling, which is the goal of ML. While there has been a growing body of work applying AI to mineral processing, these efforts have been almost entirely focused on ML (e.g., improving empirical process models, predicting metallurgical outcomes from sensor data) rather than RL \cite{mccoy2019machine, bai2025artificial}. Although model-based RL (which we employ) can incorporate ML for improved process modeling, the core objective is to learn operational policies that optimize performance over time, not to generate predictive models. To date, the mineral processing literature lacks a framework for explicitly integrating uncertainty reduction with optimization---particularly one that accounts for uncertainty arising from both feedstock variability and process complexity.

In this work, we aim to show that mineral processing operations can be framed as a problem of optimization under uncertainty, and outline the features of this approach. We then develop a mathematical formulation of a simplified flotation cell that incorporates both feedstock uncertainty and process uncertainty to inform optimization over time via data collection. We use synthetic scenarios to demonstrate the capability of this framework for optimizing the operation of a flotation cell in comparison to PID and MPC approaches, particularly in cases of significant feedstock and process uncertainty. This paper serves as a demonstration of a comprehensive mathematical approach to optimizing mineral processing under uncertainty, rather than attempting to claim that this approach ``performs better'' than existing approaches. Having highlighted the potential for this approach in various synthetic scenarios, we will discuss its potential application to real-world test cases.

\section{Features of an Optimization under Uncertainty Approach}

\subsection{Framing a Mineral Process}

\begin{figure}[htp]
\centering
\includegraphics[width=0.7\textwidth]{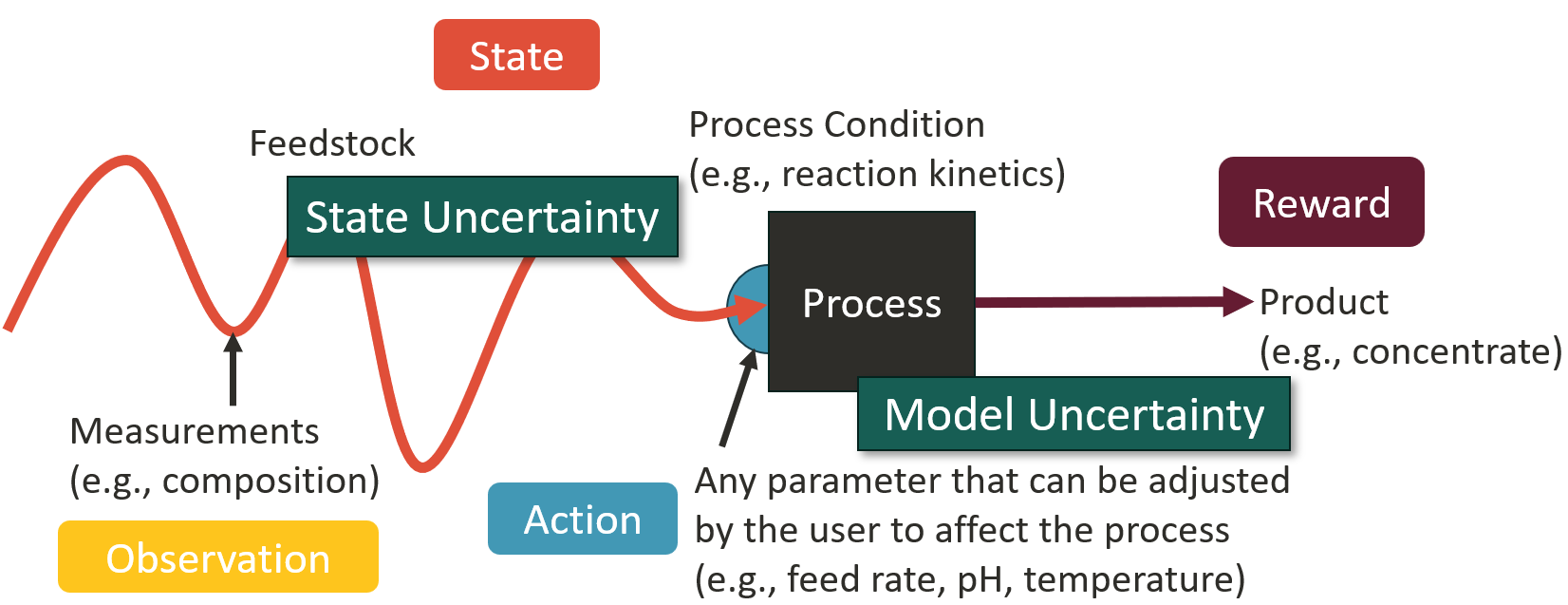}
\caption{Bare-bones framing of a mineral process (e.g., flotation) with a variable feedstock and complex process framed in terms of decision-making under uncertainty.}
\label{fig:barebones_model}
\end{figure}

Any mineral process can be viewed as a variable feed stream passing through a complex process to turn into a product. As our goal is optimization rather than to accurately describe the intricacies of a given process, we model any mineral process as a system with uncontrolled inputs, outputs, and control parameters that change how inputs translate into outputs (see Fig.~\ref{fig:barebones_model}).

To put this in the language of decision-making under uncertainty, the inputs and the process dynamics can be conceptualized as the \emph{states} of the system, control parameters as \emph{actions}, the outputs as the \emph{reward}, and any measurements taken to better ascertain the conditions of the process as \emph{observations}.

In this framework, the key uncertainties, feedstock variability and process complexity, can be classified as \emph{state uncertainty} and \emph{model uncertainty}, respectively (labeled with teal boxes in Fig.~\ref{fig:barebones_model}). Feedstock variability makes the feedstock composition and mineralogy, or the \emph{state}, uncertain, since we cannot measure every aspect of the feedstock at every point in time and space. Process complexity means we cannot know exactly how inputs (\emph{states} and \emph{actions}) translate into the output (\emph{reward}), so the model we use to describe this causal relationship is uncertain.

\subsection{The Mathematical Framework}

The principal mathematical framework for decision-making under uncertainty problems is called a Partially Observable Markov Decision Process (POMDP). A POMDP models a sequence of actions in the real world at the same time as information is gathered. This approach is now common in AI applications such as aircraft collision avoidance, self-driving cars, and robotics, and is similar to the AI used in chess and other games \cite{xiang2021recent1, xiang2021recent2}. 

Mathematically, a POMDP is defined by a tuple $\langle S, A, O, T, R, Z, \gamma \rangle$, where $S$ is the state space, $A$ is the action space, $O$ is the observation space, $T$ is the transition function, $R$ is the reward function, $Z$ is the observation function, and $\gamma$ is the discount factor \cite{arief2025managing}. The POMDP framework builds upon the Markov Decision Process (MDP) framework, a broad approach to sequential decision making in stochastic environments that forms the basis for most of reinforcement learning. In all MDPs, an \emph{agent} makes decisions (referred to as \emph{actions}) at discrete timesteps, which influence how the system transitions from one \emph{state} to the next. The current state of the system and action taken at a given timestep result in a \emph{reward} that the agent receives, which represents the optimization objective. The way that actions are chosen based on the current state or \emph{observations} of the system is called a \emph{policy}. An \emph{intelligent agent} decides a policy by learning from interacting with the system over time \cite{kochenderfer2022algorithms}.

The POMDP framework assumes that the true state of the system cannot be known. Instead, the agent holds a \emph{belief} over possible states, which is represented as a probability distribution informed by indirect, incomplete, and/or noisy observations. More information about MDPs and POMDPs can be found in the textbook \textit{Algorithms for Decision Making} \cite{kochenderfer2022algorithms}.

\begin{figure}[htp]
\centering
\includegraphics[width=0.9\textwidth]{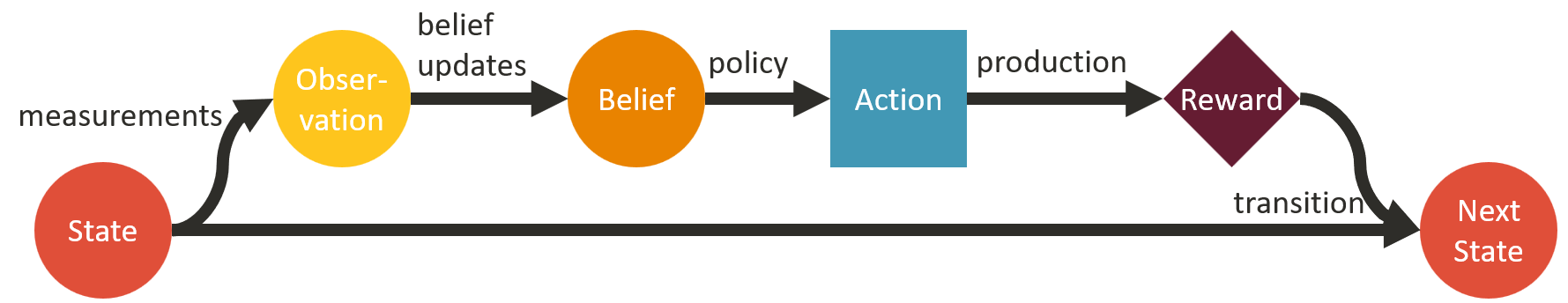}
\caption{Simplified diagram depicting the key components of a POMDP and how they progress at each timestep.}
\label{fig:pomdp}
\end{figure}

The diagram in Fig.~\ref{fig:pomdp} shows the key components of a POMDP at a given timestep, revealing its sequential nature. The labels given in Fig.~\ref{fig:barebones_model} for a generic mineral process map directly onto this framework.

\subsection{Belief: A Stochastic Representation of an Uncertain State}
\label{sec:belief}

The belief is the likelihood of being in a given state, and is typically represented by a probability distribution over states. Although the belief is not technically part of the POMDP formulation itself, it is a crucial component of decision-making under uncertainty and how POMDP solvers navigate problems. The AI literature uses belief instead of probability as a nomenclature to identify uncertainty.

\begin{figure}[htp]
\centering
\includegraphics[width=0.7\textwidth]{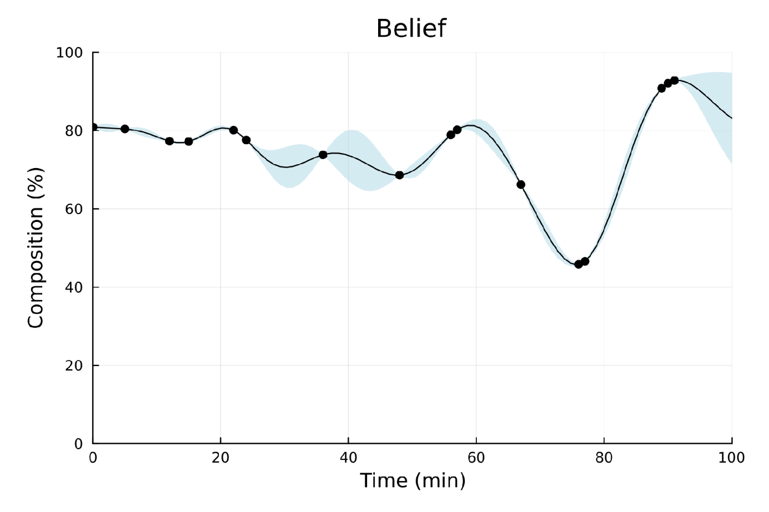}
\caption{An example of a belief at the end of a simulation, with occasional measurements (black dots) informing the belief and its associated uncertainty (blue regions).}
\label{fig:belief_example}
\end{figure}

The intelligent agent forms a belief of the state based on measurements that it receives. As shown in Fig.~\ref{fig:belief_example}, the belief evolves over time as more measurements are collected. Uncertainty, represented by variance values, is captured in the belief's nature as a probability distribution rather than a discrete quantity or set of quantities.

\subsection{Reward: The Optimization Objective}

\begin{figure}[htp]
\centering
\includegraphics[width=0.7\textwidth]{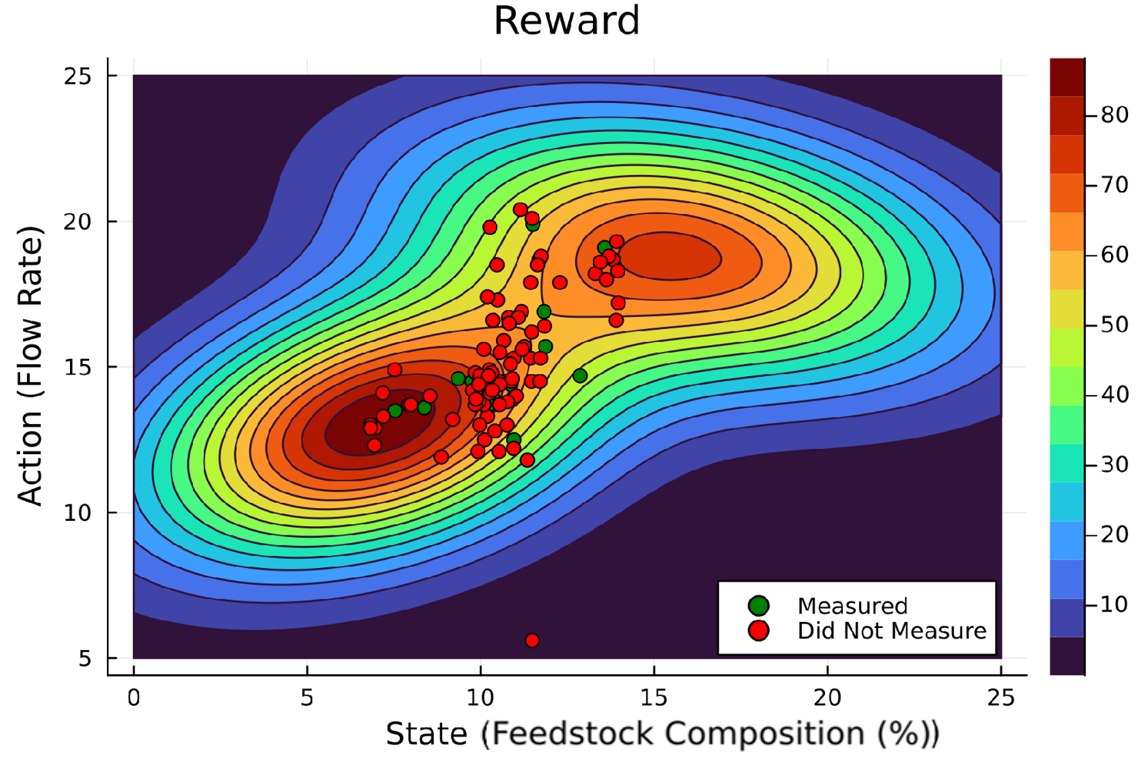}
\caption{An example of a reward surface (reward as a function of state and action) showing how the intelligent agent balances a cost associated with taking measurements with the cost of choosing a poor action. Actions are taken as the state fluctuates in time, with red dots representing times when measurements were not taken, and green dots representing times when measurements were taken.}
\label{fig:reward_example}
\end{figure}

The reward function describes the optimization objective. Value judgments from experts are necessary to define what the ultimate goal of the optimization should be. As shown in Fig.~\ref{fig:reward_example}, the reward captures the inherent tradeoff between the cost of measuring and the cost of not having information.

\section{Mathematical Formulation of a Flotation Cell}
\label{sec:flotation_pomdp}

\begin{figure}[htp]
\centering
\includegraphics[width=0.8\textwidth]{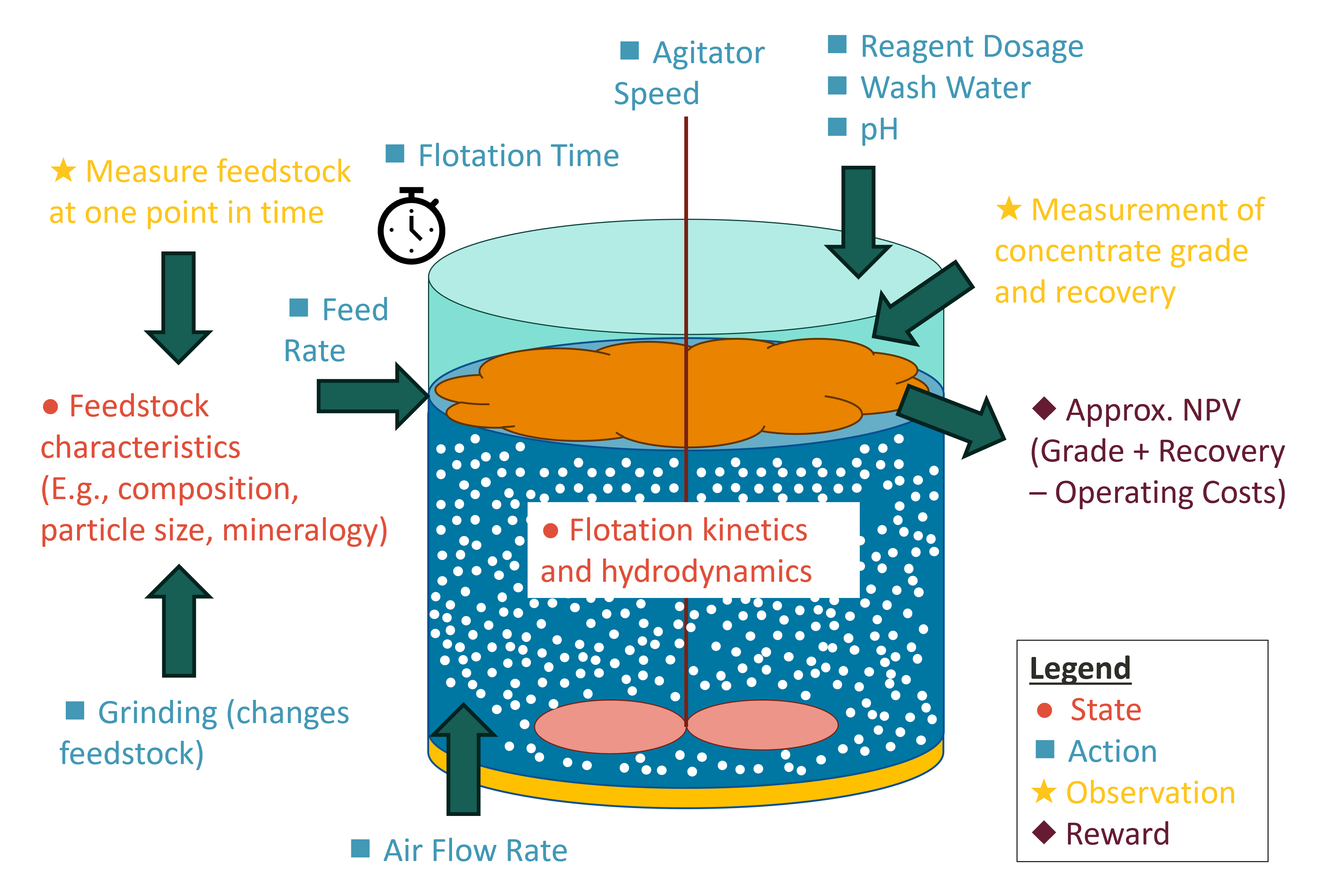}
\caption{A simple POMDP formulation of a flotation cell. Additional possible state variables and control parameters beyond the scope of the formulation in this paper are included as examples.}
\label{fig:flotation_pomdp}
\end{figure}

Now that we have established the general concept of applying an optimization-under-uncertainty approach to mineral processing, we can formulate the operation of a flotation cell as a POMDP. A simple formulation is shown in Fig.~\ref{fig:flotation_pomdp}, labeling a few components of a flotation cell under this framework.

For now, we represent the flotation cell as a batch process, where one batch is processed at each timestep. This allows for straightforward experimental validation at bench scale. The formulation can be adapted for a continuous process, as is typical at the industrial scale.

The values used throughout our flotation cell formulation and implementation are meant to roughly reflect typical values for phosphate flotation as an example.

\subsection{State}
The state is represented by the following variables:
\begin{itemize}
    \item Feedstock composition $c$
    \item Concentrate recovery $r$
    \item Concentrate grade $g$
    \item Timestep $T$
\end{itemize}
For simplicity and clarity of presentation, we represent the feedstock characteristics with one variable, the average composition for a given batch. The state includes time mostly as a technicality, since the transition function (which is a function of the state and action) depends on time.

\subsection{Actions}
Actions are control parameters that can be adjusted to change the operating conditions of the flotation cell, as well as the decision to make measurements. For simplicity and clarity of presentation, we choose two control parameters as our action set.
\begin{itemize}
    \item Flotation time $t$ (min)
    \item Air flow rate $f$ (L/hr)
    \item Measure feedstock
\end{itemize}

For the model uncertainty tests in Sec.~\ref{sec:model_uncertainty}, the \textsc{Measure feedstock} action is always set to \textsc{true}. \textsc{Measure feedstock} becomes a choice between \textsc{true} and \textsc{false} for the feedstock uncertainty tests in Sec.~\ref{sec:state_uncertainty}.

\subsection{Transition Function}
The transition function describes how the current state transitions to the next state as a function of the current state and action. In other words, it corresponds to a forward model that describes how the system changes in time.

The transition model is described by a combination of the following:
\begin{itemize}
    \item the simple kinetic model (see Sec.~\ref{sec:flotation_model})
    \item a stochastic representation (e.g., Gaussian process) of the feedstock composition fluctuating in time
    \item stochastic representations (e.g., Gaussian process) of the errors between the kinetic model and the true grade and recovery
\end{itemize}

Note that in this formulation, transition probabilities are only dependent on the state, not the action. Also, we consider actions to have a deterministic effect. In other words, if we were to know the state, then choosing an action would deterministically result in a given reward. Transition uncertainty can be introduced by making actions stochastic (in other words, imprecise).

\subsection{Observations}
The observations are:
\begin{itemize}
    \item Average feedstock composition (can be null)
    \item True recovery and grade
\end{itemize}

For the model uncertainty tests in Sec.~\ref{sec:model_uncertainty}, full observations of the state are received at every timestep, so the implementation technically reduces to an MDP (no state uncertainty, only transition uncertainty). For the feedstock uncertainty tests in Sec.~\ref{sec:model_uncertainty}, observations of feedstock composition are only received when the \textsc{measure feedstock} action is taken. Otherwise, no information about the feedstock is collected at that timestep.

\subsection{Observation Function}
The observation function is the likelihood of an observation given a state. For the scope of this paper, since the goal is more to demonstrate the approach, the observation function is just a delta function (i.e., an exact observation). The observation returned at each timestep is simply the true state.

\subsection{Reward}
\label{sec:reward}
Here, we consider the reward to be an approximation of the net present value (NPV) of the process. We use the Moroccan phosphate industry (i.e., the OCP Group) as an example. The specific formula for the reward defined in Eq.~\ref{eq:reward} uses back-of-the-envelope estimates for the current production of phosphate concentrate as a function of recovery, the price as a function of grade, and the operating costs \cite{usgs2024phosphate, worldbank2025pink}. The operating cost formula in Eq.~\ref{eq:opex} is not intended to reflect realistic operating costs, but rather is designed to create a global optimum from the tradeoff between grade, recovery, and operating costs. Examples of the reward at a fixed feedstock composition depicted in Fig.~\ref{fig:reward} exhibit this tradeoff.

\begin{align}
    reward &= \frac{500g \text{ [\$/t]} \cdot 35 r \text{ [Mt/yr]}}{100 \text{ [timestep/yr]}} - OPEX \text{ [\$M/timestep]} \label{eq:reward} \\
    OPEX &= \frac{1}{2} t + \frac{1}{50} f \text{ [\$M/timestep]} \label{eq:opex}
\end{align}

\section{Simple Flotation Model}
\label{sec:flotation_model}

We lay out a simple flotation model describe the system to be optimized. This model of the system corresponds to the \emph{transition function} in a POMDP formulation.

In flotation, recovery and grade are the key performance metrics, and there is a natural tradeoff between the two. (As recovery approaches 100\%, concentrate grade approaches feed grade, and as concentrate grade approaches 100\%, recovery approaches 0\%. A grade-recovery curve then essentially forms a Pareto front.)

We model the black box mechanical flotation cell with empirically-inspired equations. The goal is to capture broad relationships between the inputs, control parameters, and outputs, rather than to be accurate.

The instantaneous recovery $r$ and the instantaneous concentrate grade $g$ are both reported in percentage (i.e., ranging from 0 to 100),
\begin{align}
    r(k, t, f) &= 100 \frac{kt}{1+kt} \frac{f}{f+10} \label{eq:recovery} \\
    g(c, k, t, f) &= c \left[ 1 + \left( 1 - \frac{c}{42.2}\right) \left( 1 - \frac{\exp{\left(-kt/10\right)}}{1 + \exp{\left(4 - 0.04f\right)}} \right) \right] \label{eq:grade} 
\end{align}
where $c$ is the feedstock composition (i.e., feed grade) in percentage, $k$ is the flotation rate constant in min$^{-1}$ (set to 1), $t$ is the flotation time in minutes, and $f$ is the air flow rate in L/hr.

\begin{figure}[p]
\centering
\begin{subfigure}[c]{0.3\textwidth}
  \includegraphics[width=\textwidth]{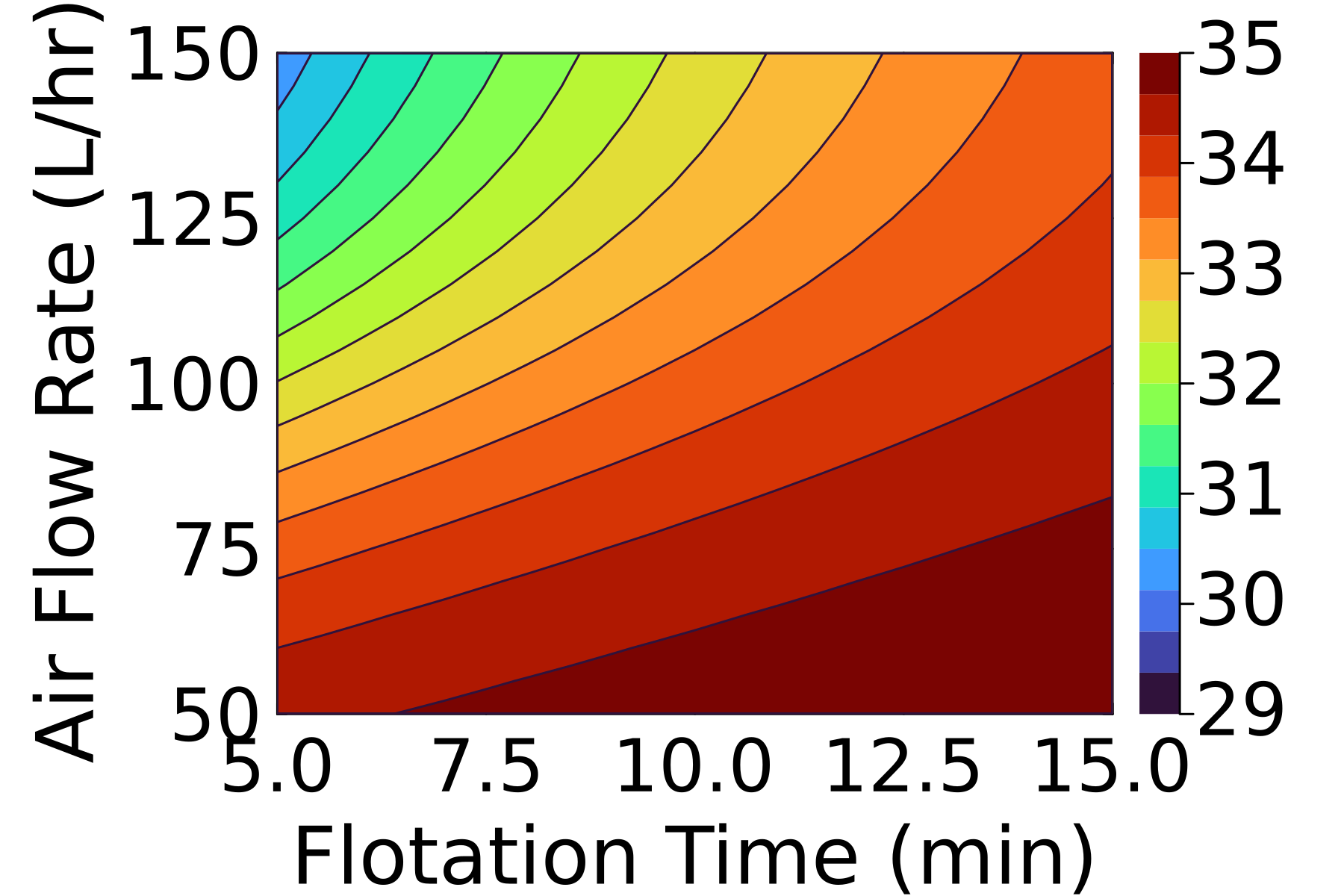}
  \caption{Kinetic model}
  \label{fig:grade_kinetic}
\end{subfigure}
\begin{subfigure}[c]{0.03\textwidth}
  \textbf{+}
\end{subfigure}
\begin{subfigure}[c]{0.3\textwidth}
  \includegraphics[width=\textwidth]{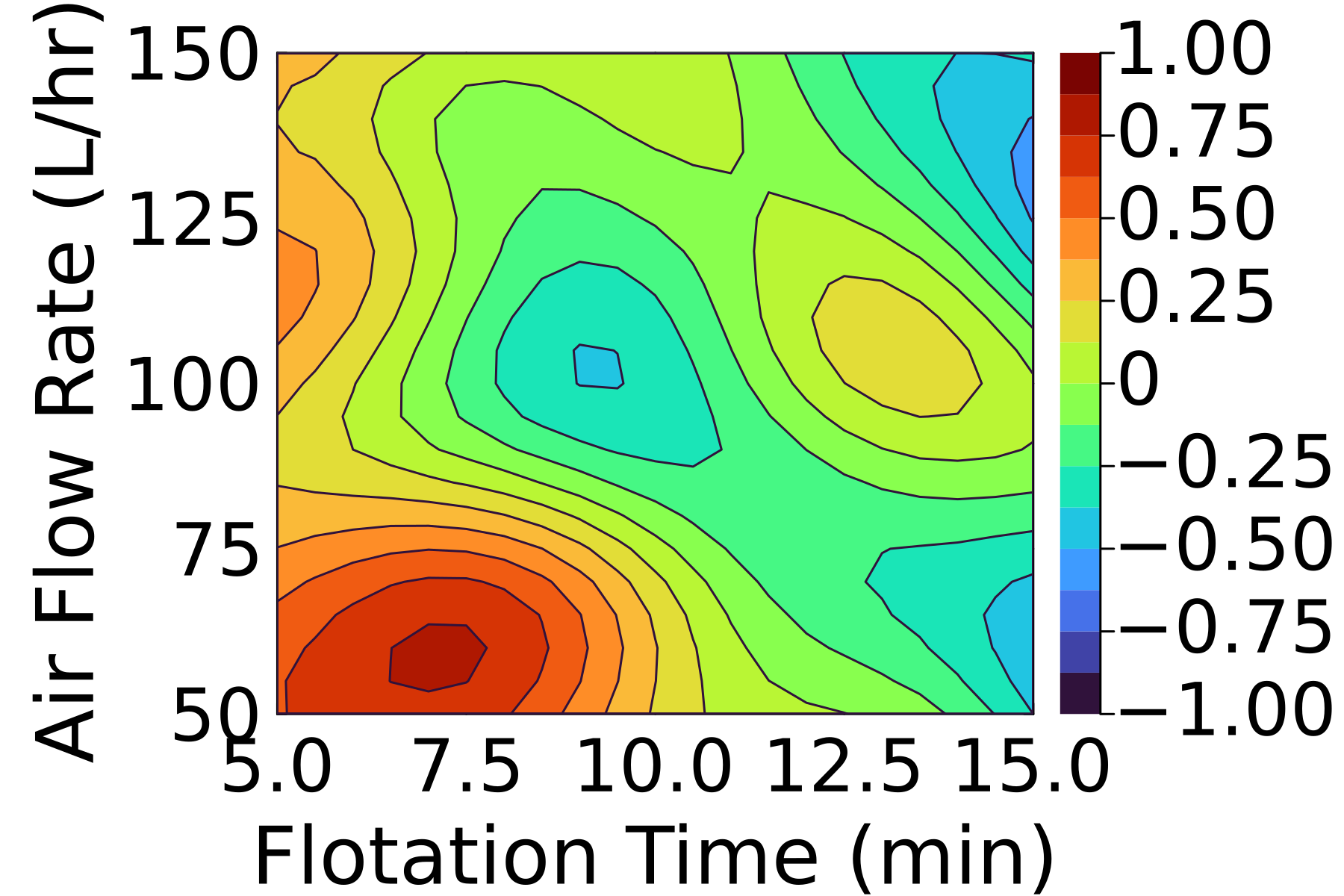}
  \caption{Error function}
  \label{fig:grade_error}
\end{subfigure}
\begin{subfigure}[c]{0.03\textwidth}
  \textbf{=}
\end{subfigure}
\begin{subfigure}[c]{0.3\textwidth}
  \includegraphics[width=\textwidth]{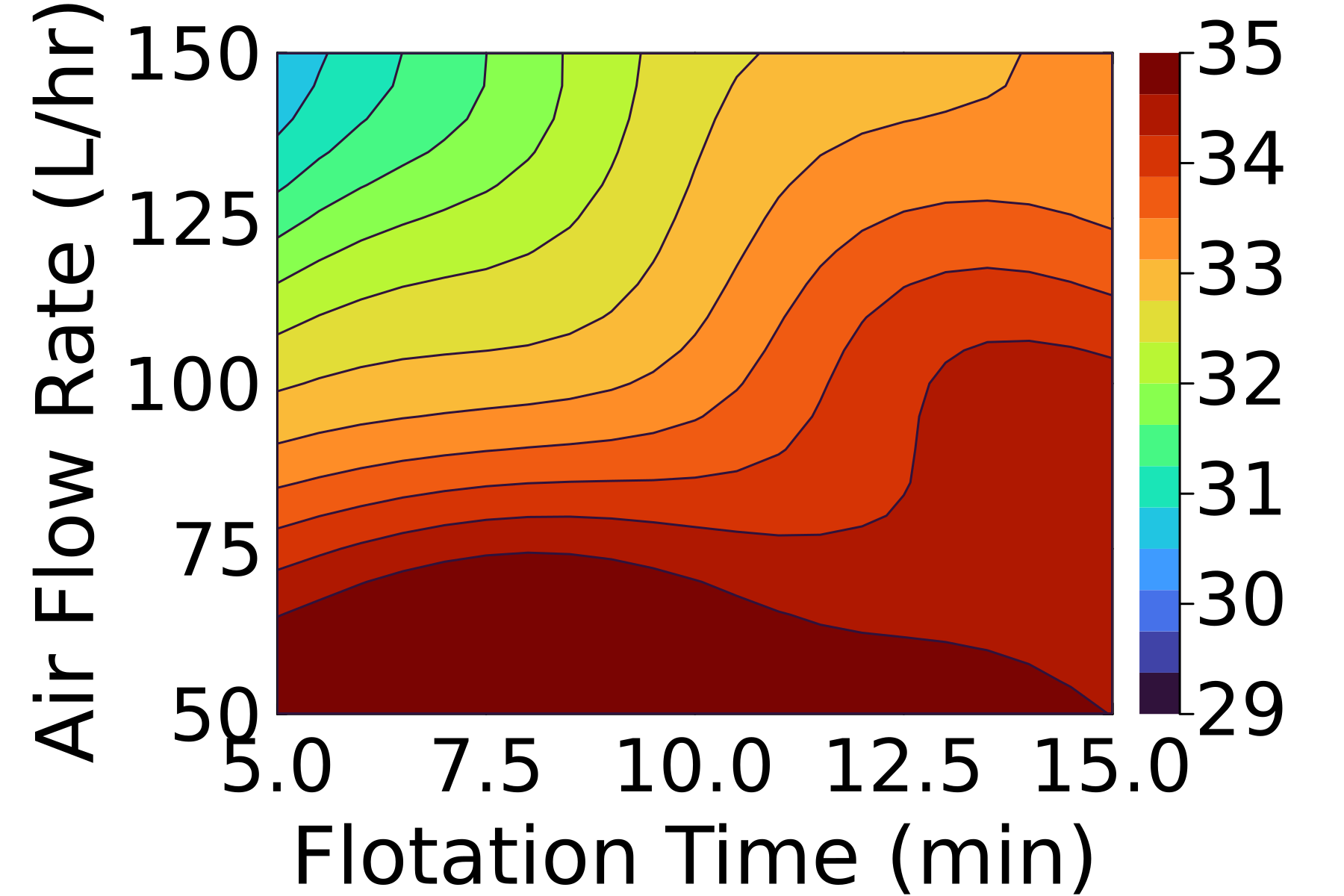}
  \caption{True grade}
  \label{fig:grade_true}
\end{subfigure} 
\caption{An example of the grade (\%) as a function of the actions (air flow rate and flotation time) at a fixed feedstock composition.}
\label{fig:grade}
\end{figure}

\begin{figure}[htp]
\centering
\begin{subfigure}[c]{0.3\textwidth}
  \includegraphics[width=\textwidth]{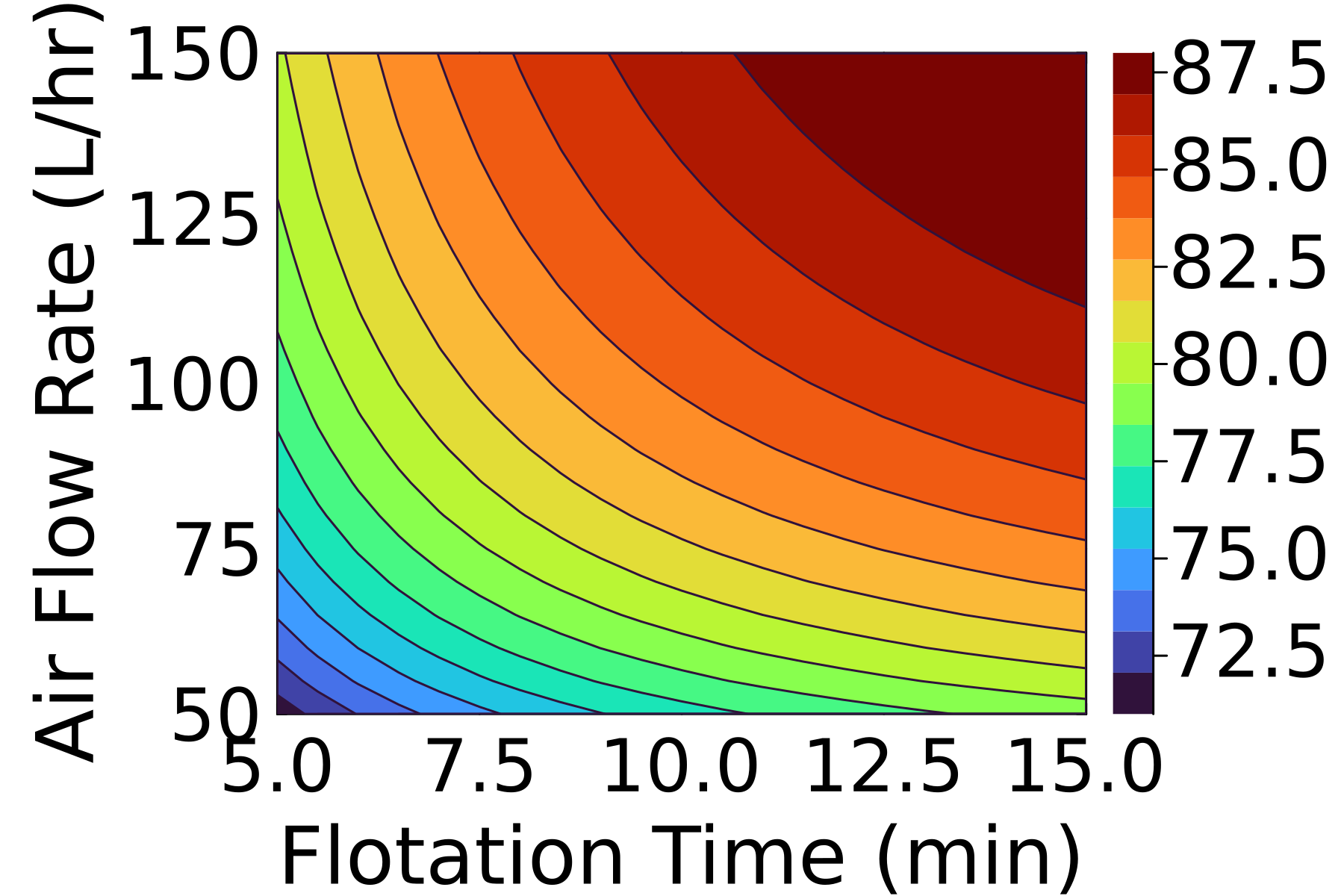}
  \caption{Kinetic model}
  \label{fig:recovery_kinetic}
\end{subfigure}
\begin{subfigure}[c]{0.03\textwidth}
  \textbf{+}
\end{subfigure}
\begin{subfigure}[c]{0.3\textwidth}
  \includegraphics[width=\textwidth]{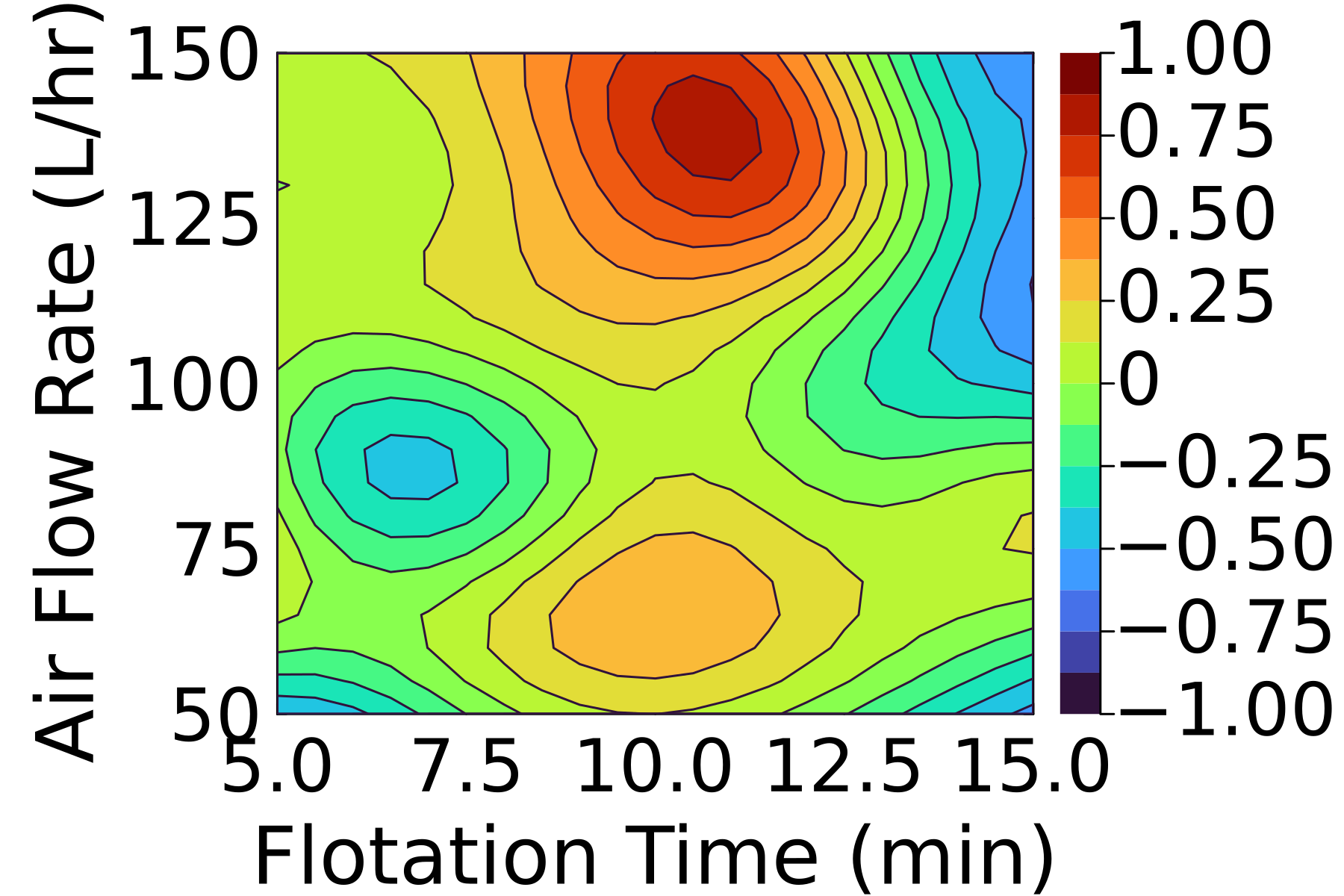}
  \caption{Error function}
  \label{fig:recovery_error}
\end{subfigure}
\begin{subfigure}[c]{0.03\textwidth}
  \textbf{=}
\end{subfigure}
\begin{subfigure}[c]{0.3\textwidth}
  \includegraphics[width=\textwidth]{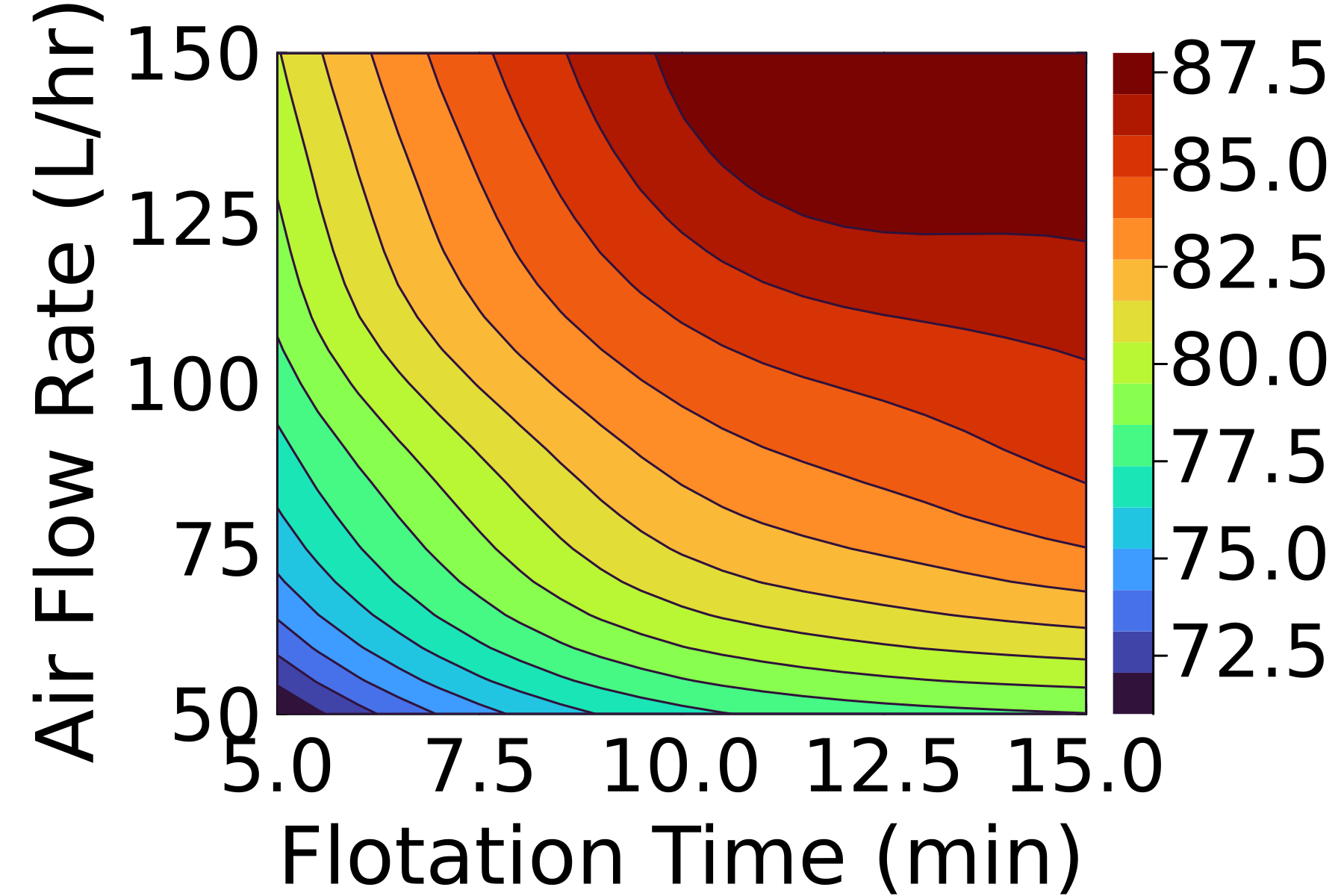}
  \caption{True recovery}
  \label{fig:recovery_true}
\end{subfigure} 
\centering
\caption{An example of the recovery (\%) as a function of actions (air flow rate and flotation time) at a fixed feedstock composition.}
\label{fig:recovery}
\end{figure}

The exact equations are mostly arbitrary, with numbers that very roughly correspond to phosphate flotation. The grade equation in particular is set up to refer to P$_2$O$_5$ grade. They are designed to highlight the tradeoffs present in flotation cells, as can be seen in the plots of this simple kinetic model in Figs.~\ref{fig:grade_kinetic} \& \ref{fig:recovery_kinetic}.

\begin{figure}[htp]
\centering
\begin{subfigure}[c]{0.48\textwidth}
  \includegraphics[width=\textwidth]{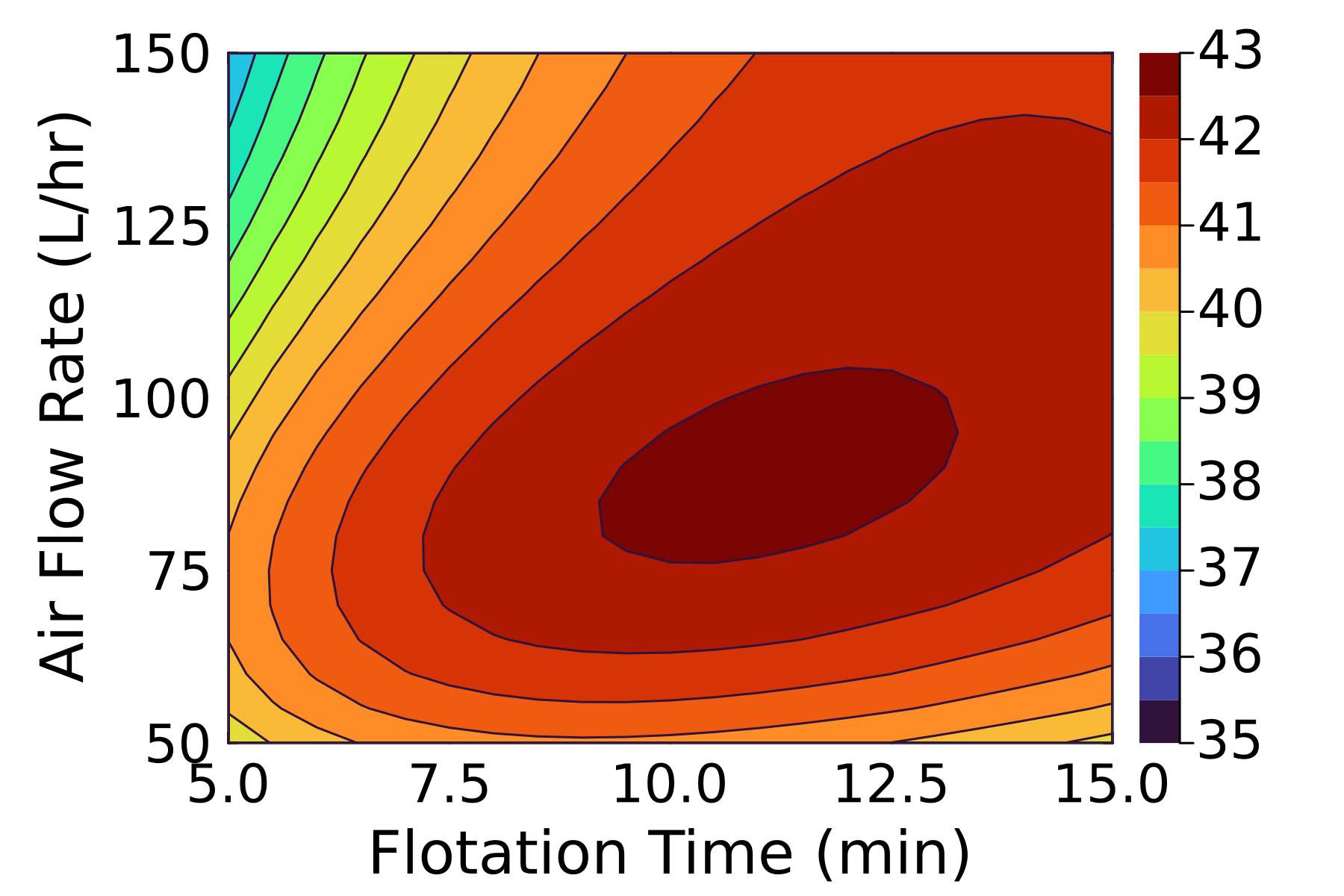}
  \caption{Kinetic model}
  \label{fig:reward_kinetic}
\end{subfigure}
\begin{subfigure}[c]{0.48\textwidth}
  \includegraphics[width=\textwidth]{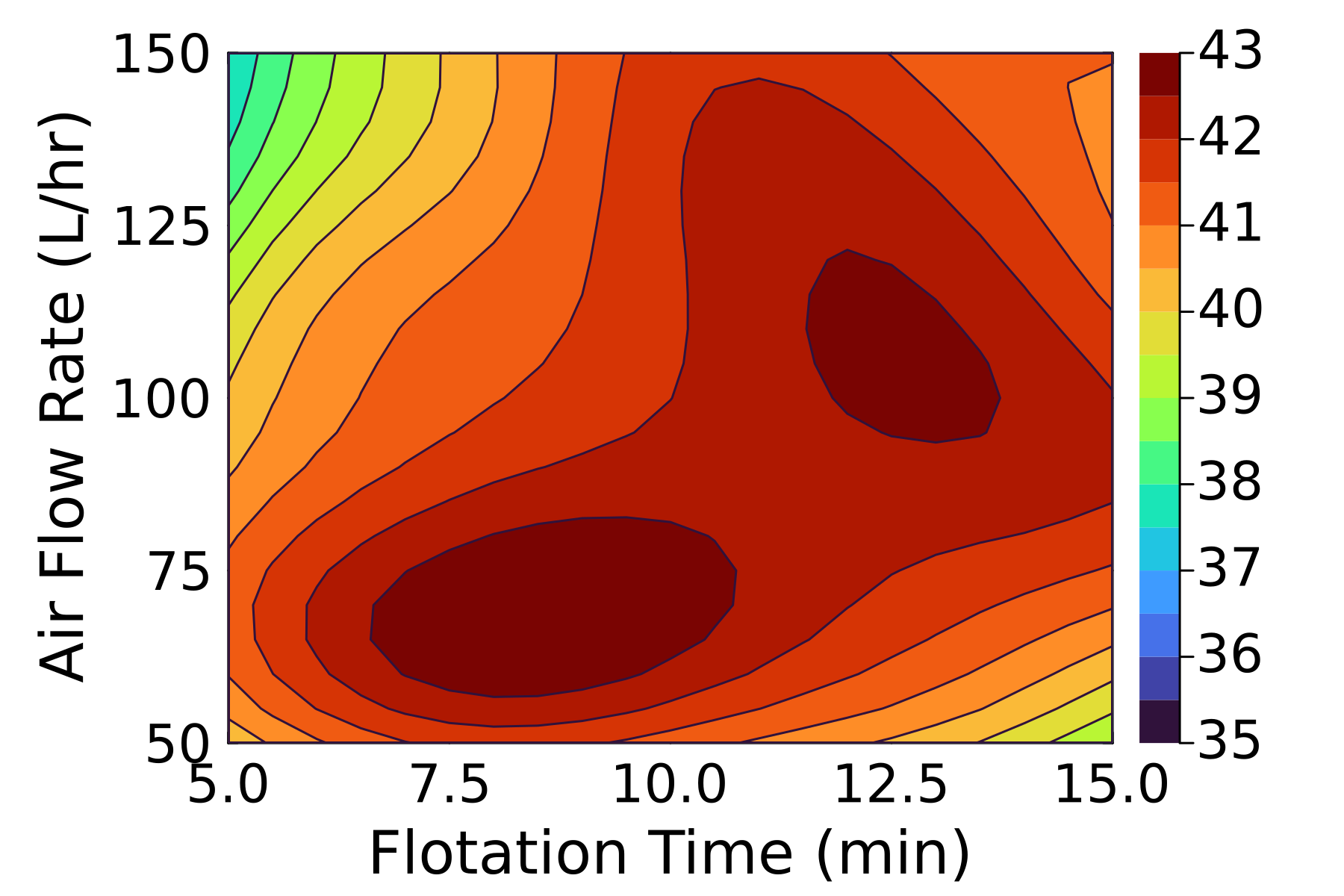}
  \caption{True reward}
  \label{fig:reward_true}
\end{subfigure} 
\caption{Examples of the reward (NPV) as a function of actions (air flow rate and flotation time) at a fixed feedstock composition.}
\label{fig:reward}
\end{figure}

We do not expect our simple kinetic model to accurately capture flotation dynamics, but rather to serve as a reasonable first guess or \emph{prior}. We can capture all inaccuracies in our model as well as any inaccuracy in any measurement of the true grade and recovery in an \emph{error function}. Then, we can represent the true grade or recovery as the sum of the kinetic model and some stochastic error function, as depicted in Figs.~\ref{fig:grade} \& \ref{fig:recovery}. The true grade and recovery result in a true reward function as well, shown in Fig.~\ref{fig:reward}.

For the synthetic cases in this paper, the ``true'' error function is generated stochastically to produce a ground truth grade and recovery that represent ``reality''. Just like in real life, this ground truth is unknown to the intelligent agent seeking to optimize the flotation cell, but can be explored through measurements.

\section{Belief Update}
\label{sec:belief_update}

As introduced in Sec.~\ref{sec:belief}, the intelligent agent updates the belief to learn the true grade and recovery and improve upon our prior model of the system.

In the flotation problem, the belief is represented by:
\begin{itemize}
    \item Gaussian process of feedstock composition
    \item Gaussian process of grade and recovery error functions
\end{itemize}

Uncertainty is represented stochastically with Gaussian processes. Actions (i.e., setting the air flow rate and flotation time) are chosen at each point in time as the feedstock composition fluctuates. The measured grade and recovery then inform the updated belief, which helps improve decision-making. The Gaussian processes in the belief are updated by sequentially refitting them to include the new data.

\begin{figure}[htp]
\centering
\begin{subfigure}[c]{0.325\textwidth}
  \includegraphics[width=\textwidth]{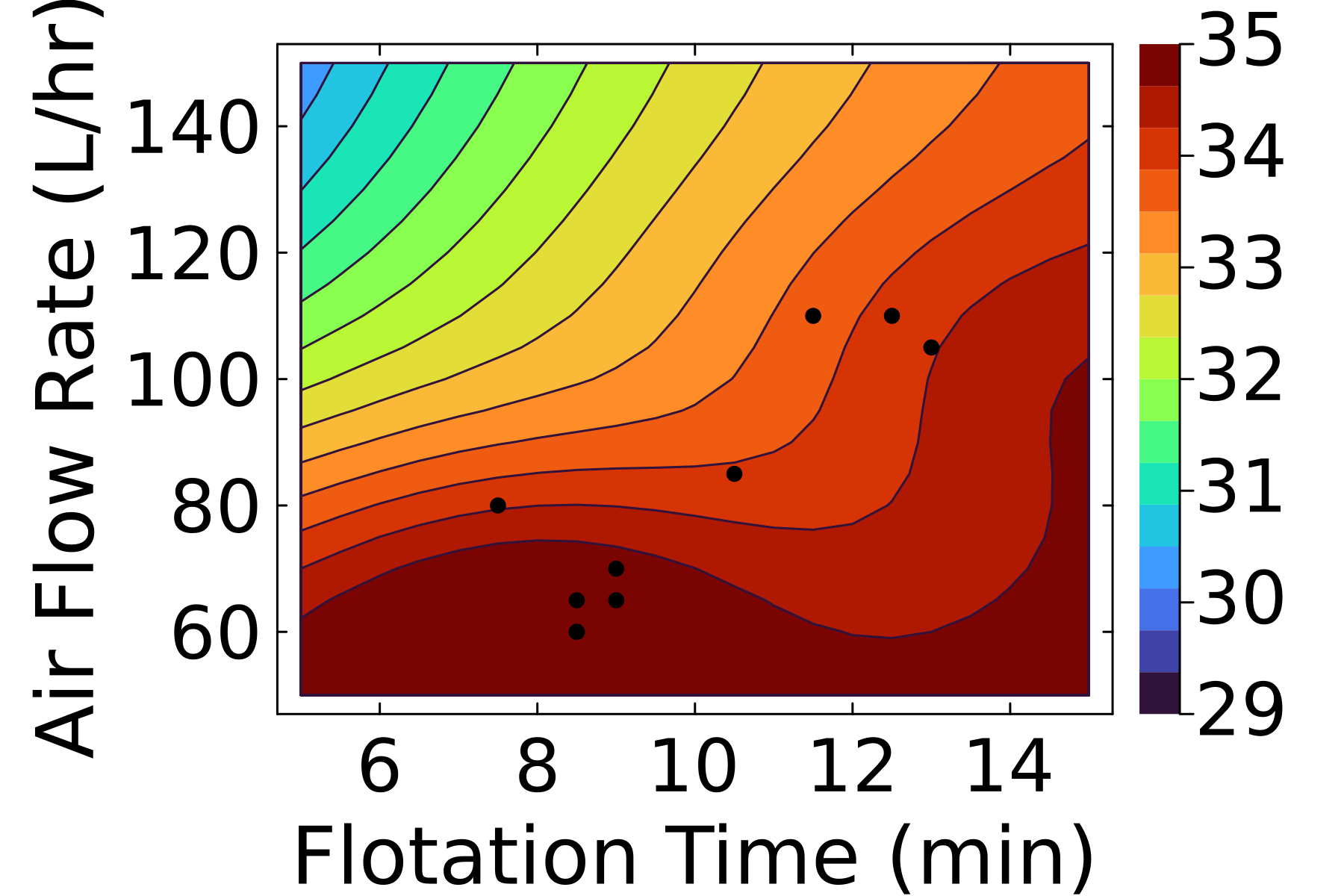}
  \caption{Grade belief (\%)}
  \label{fig:grade_belief}
\end{subfigure}
\begin{subfigure}[c]{0.325\textwidth}
  \includegraphics[width=\textwidth]{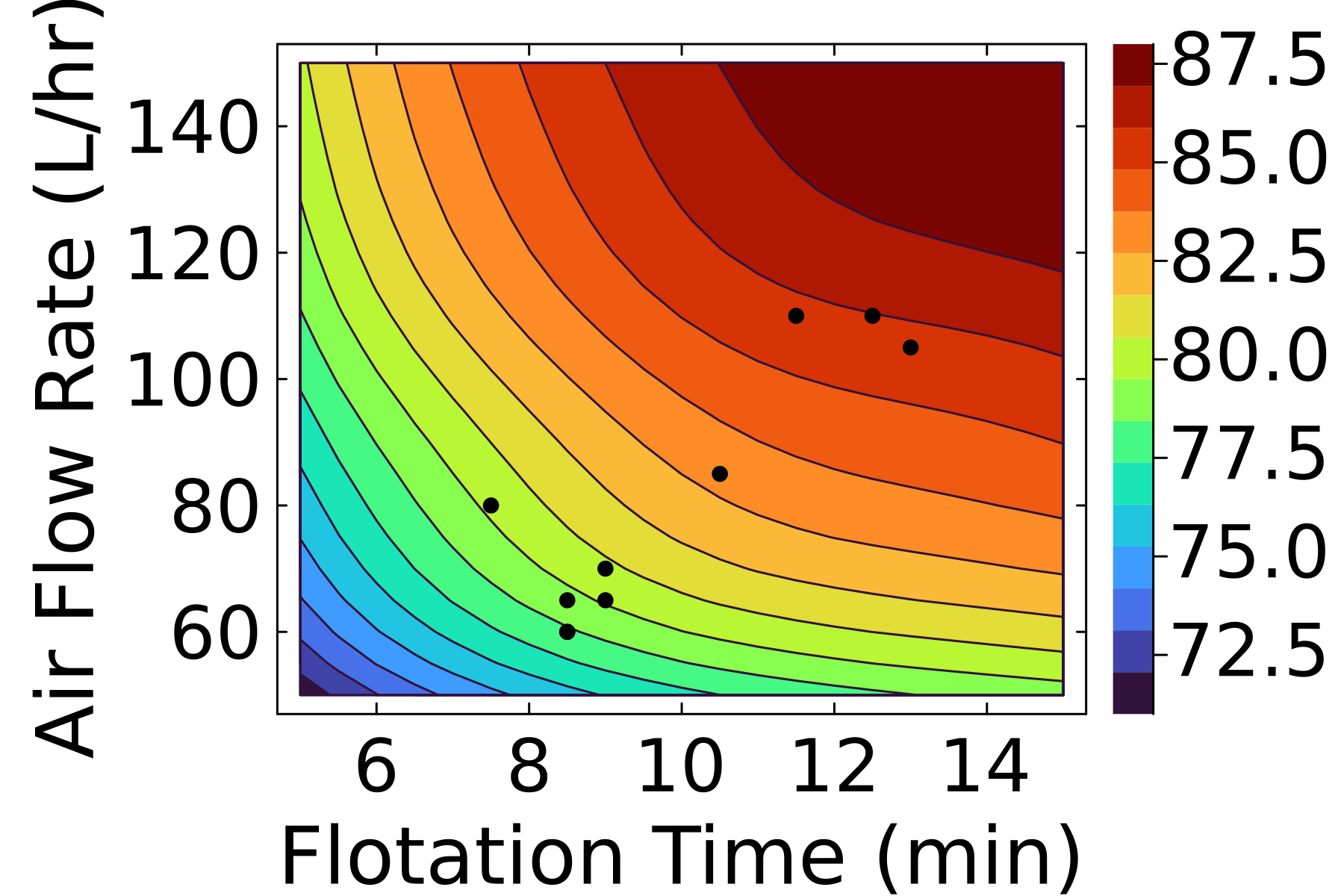}
  \caption{Recovery belief (\%)}
  \label{fig:recovery_belief}
\end{subfigure}
\begin{subfigure}[c]{0.325\textwidth}
  \includegraphics[width=\textwidth]{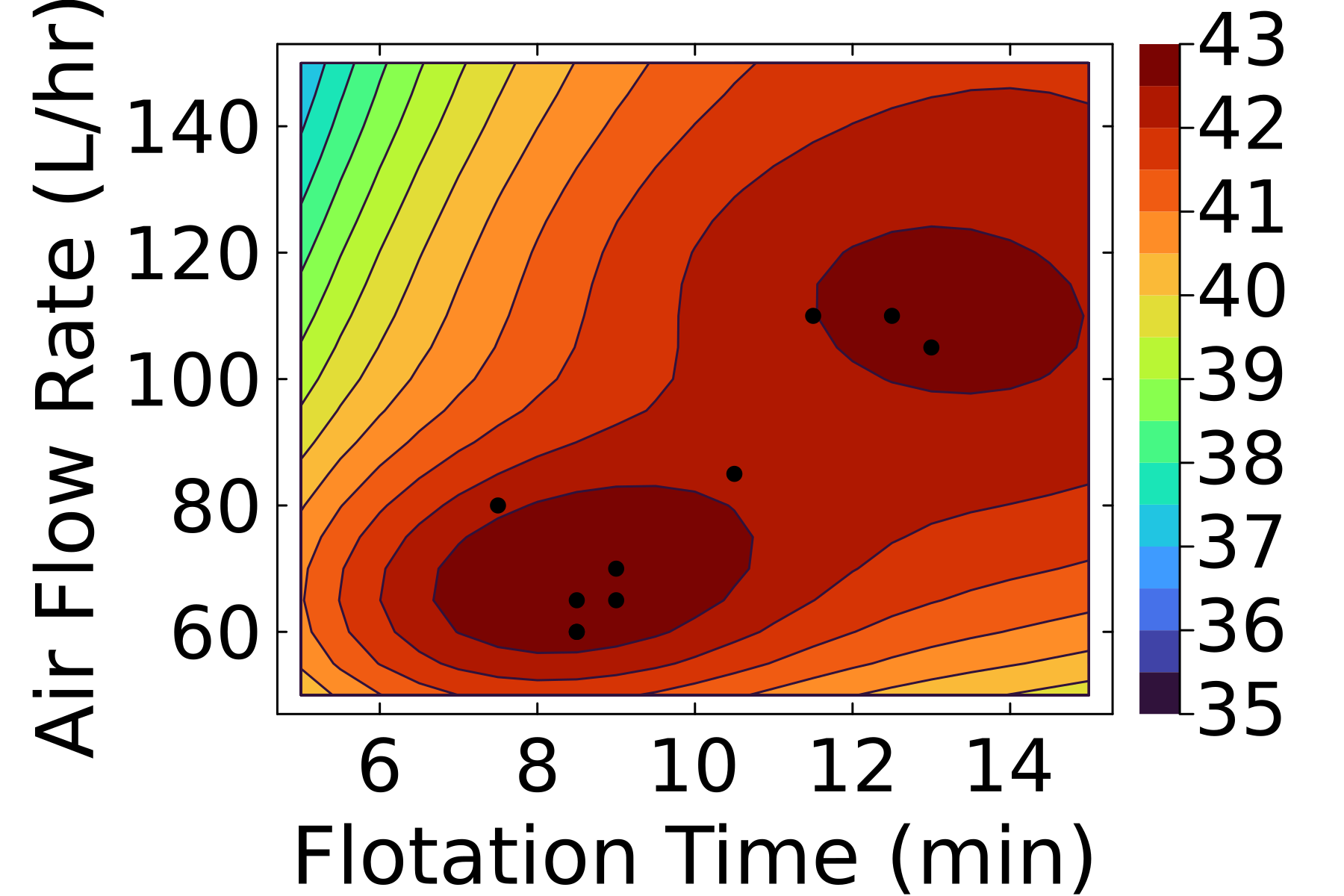}
  \caption{Reward belief (NPV)}
  \label{fig:reward_belief}
\end{subfigure} 
\centering
\caption{Beliefs represented by Gaussian processes that are progressively updated as new data is collected over time. Black dots represent collected data.}
\label{fig:belief}
\end{figure}

\section{Simulation Setup}

We investigate the extent to which the performance of different approaches is affected by feedstock (i.e, state) and process (i.e., model) uncertainty. We do so by running simulations of 100 timesteps, which represent 100 flotation batches processed over one year.

\subsection{Establishing the Baseline}

In a POMDP framework, a given approach to choosing actions based on the current belief is called a \emph{policy}. Control and optimization algorithms can be considered types of policies. As stated in Sec.~\ref{sec:intro}, two commonly-used deterministic methods, PID and MPC, are used as a frame of reference. Although MPC is typically used as a control method, we implement MPC using an optimization approach with the goal of maximizing the \emph{reward}, not just the grade and recovery, to serve as a direct comparison to the POMDP approach.

\subsection{Performance Metric}

Since we are aiming for the goal of process optimization, the reward (proxy for NPV, established in Sec.~\ref{sec:reward}) is used as the metric of comparison.

\subsection{POMDP Solver}

To solve the problem we have now formulated as a POMDP, we use a well-established online solver called Partially Observable Monte Carlo Planning (POMCP) \cite{silver2010monte}. To determine a policy, POMCP uses Monte Carlo tree search (MCTS), a common algorithm for deciding a course of action from many possible futures. The most well-known application of MCTS is in game-playing AI.

\section{Demonstration of Optimization-under-Uncertainty Approach}

\subsection{Optimization under Model Uncertainty}
\label{sec:model_uncertainty}

\begin{figure}[htp]
\centering
\begin{subfigure}[c]{0.24\textwidth}
  \includegraphics[width=\textwidth]{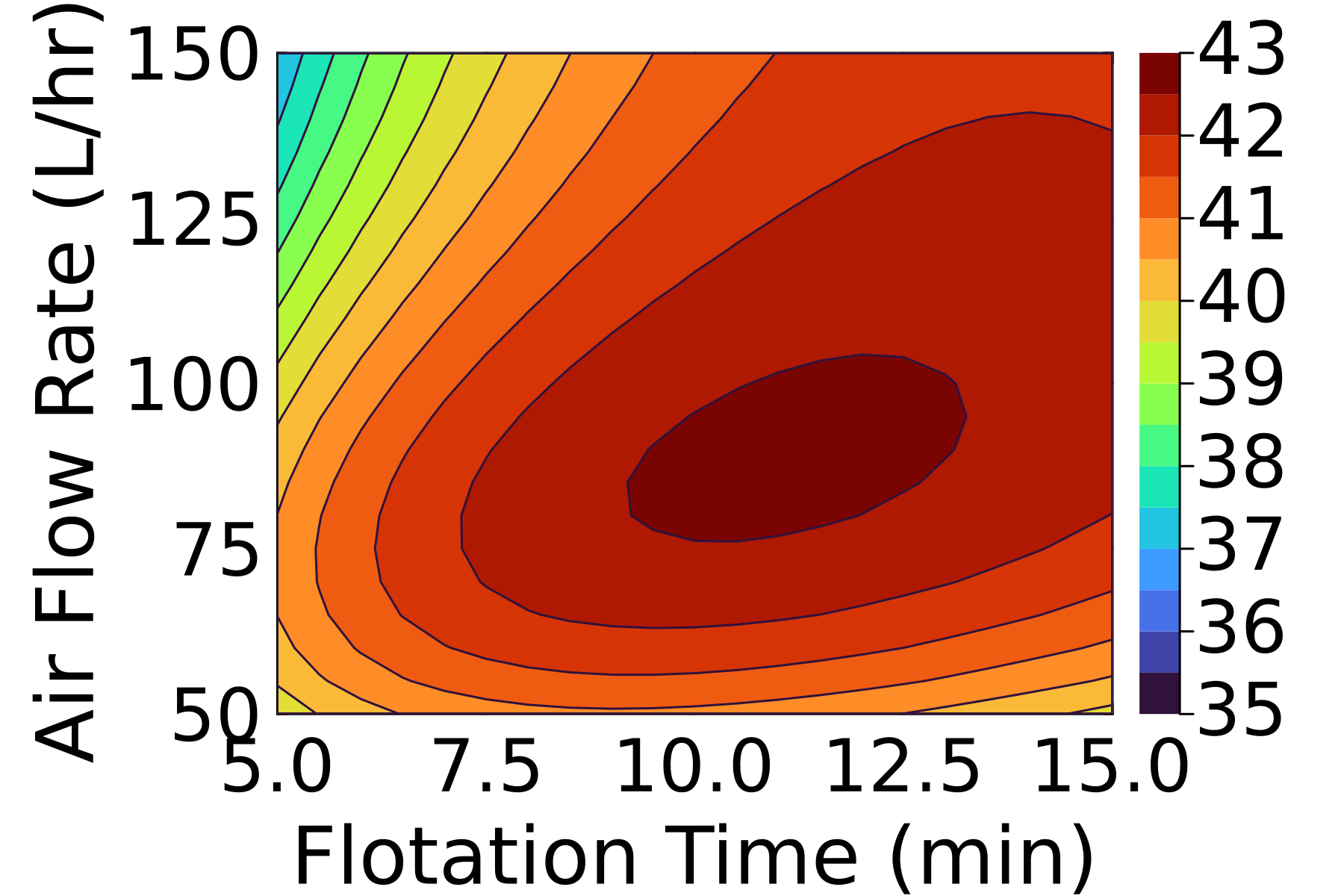}
  \caption{Kinetic model}
\end{subfigure}
\unskip\ \vrule\ 
\begin{subfigure}[c]{0.24\textwidth}
  \includegraphics[width=\textwidth]{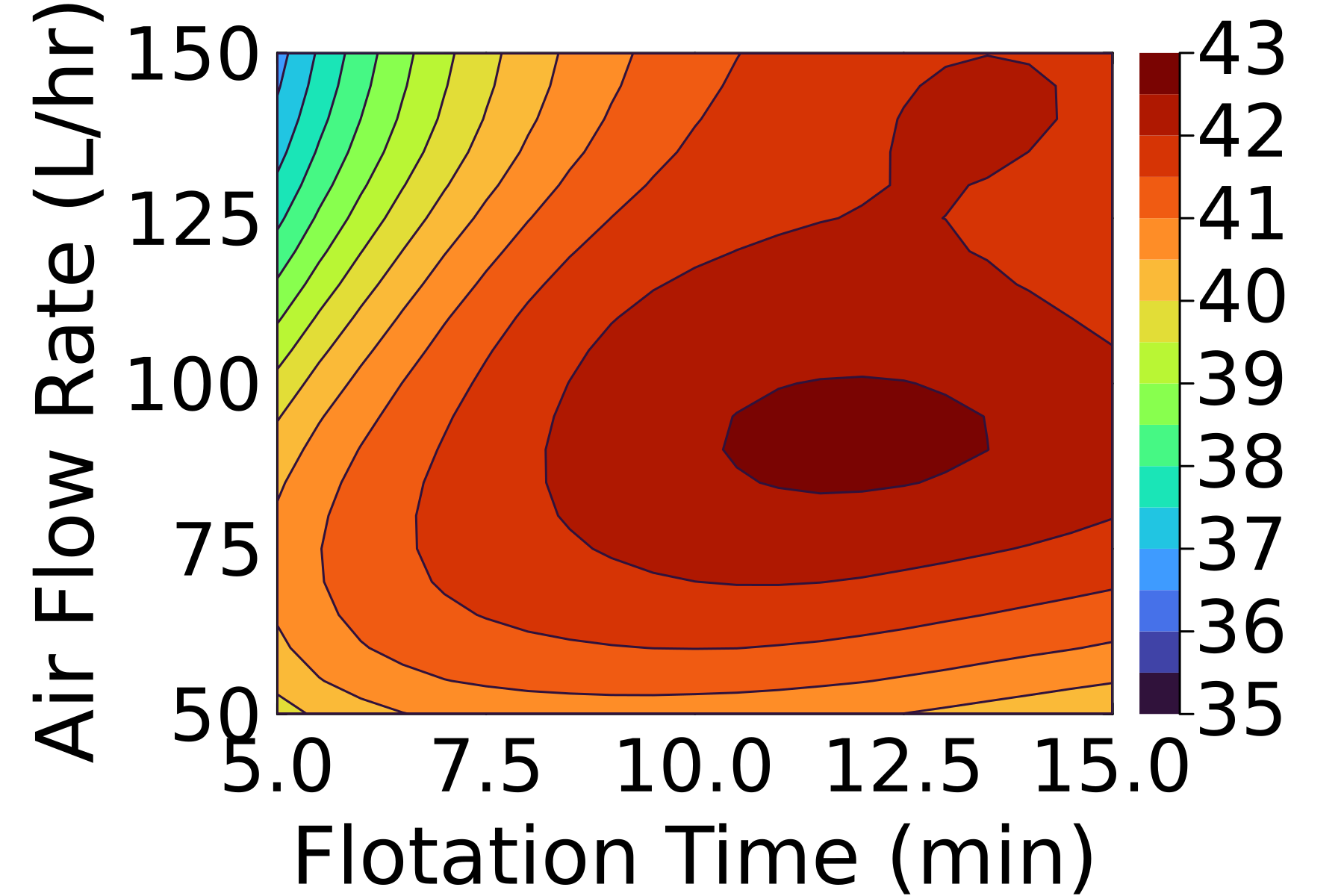}
  \caption{High accuracy}
\end{subfigure}
\begin{subfigure}[c]{0.24\textwidth}
  \includegraphics[width=\textwidth]{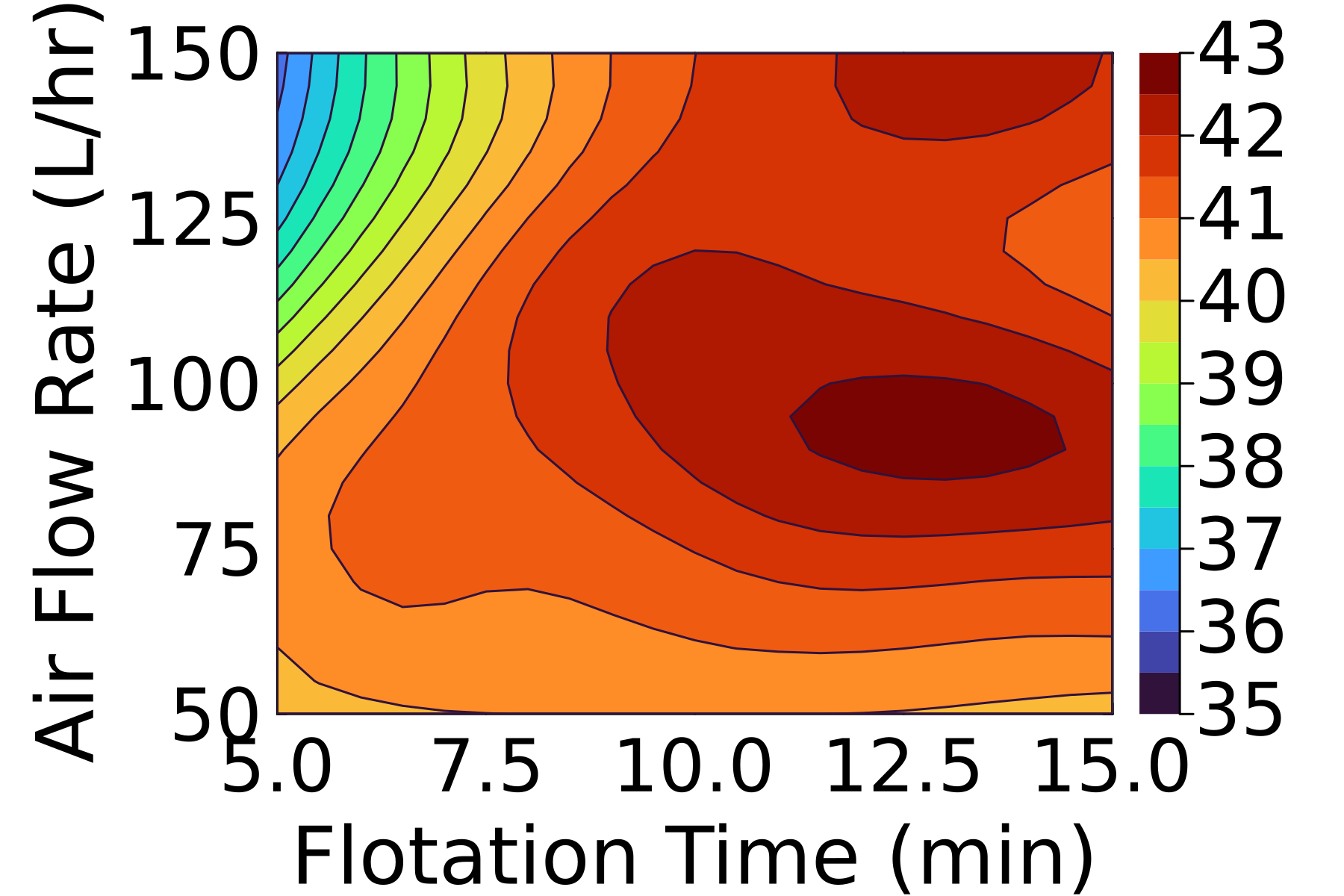}
  \caption{Medium accuracy}
\end{subfigure}
\begin{subfigure}[c]{0.24\textwidth}
  \includegraphics[width=\textwidth]{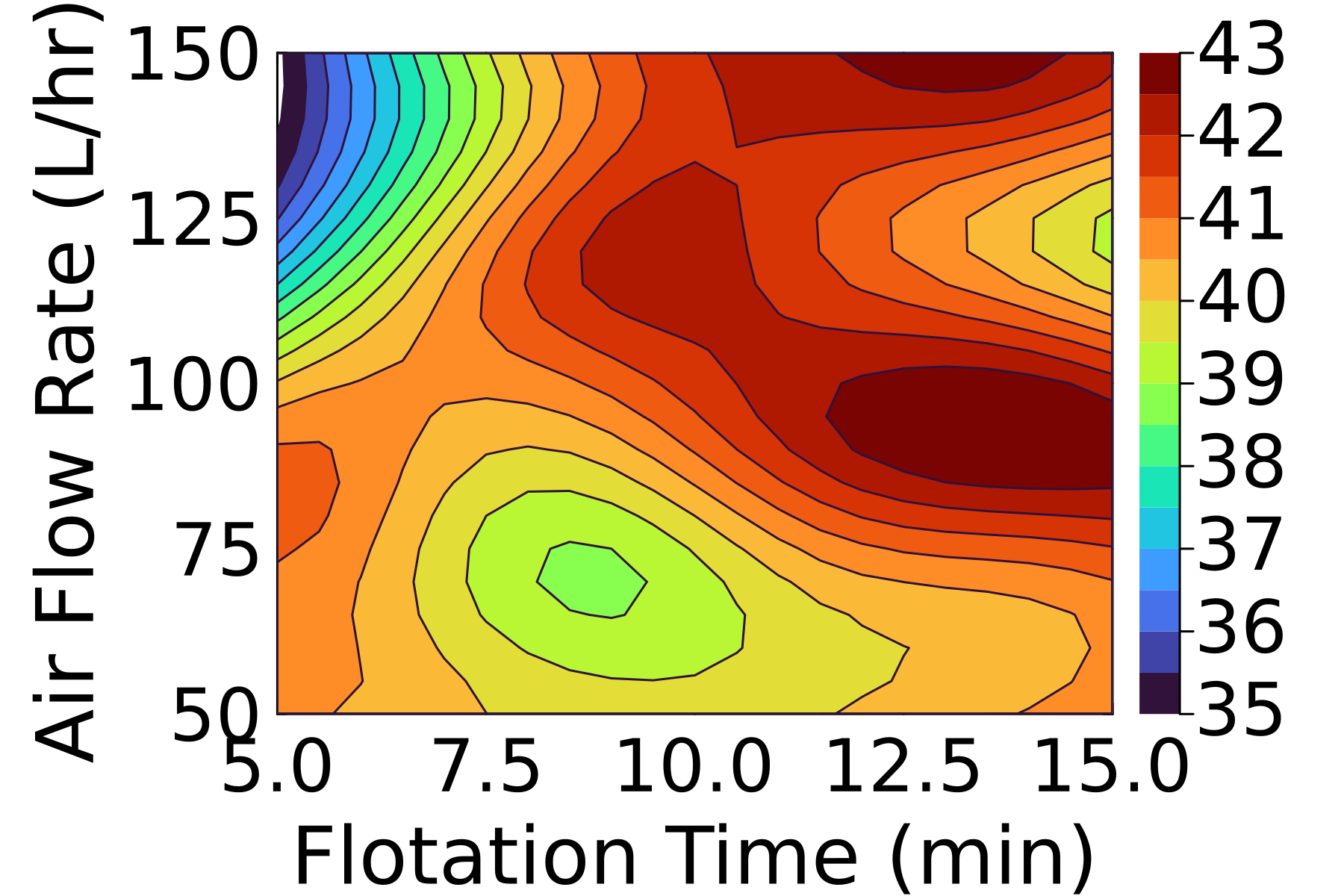}
  \caption{Low accuracy}
\end{subfigure}
\centering
\caption{Reward functions (NPV) of varying degrees of similarity to the kinetic model.}
\label{fig:model_uncertainty}
\end{figure}

To evaluate optimization under model uncertainty, we consider three cases of differing degrees of model uncertainty: where the simple flotation model has high, medium, and low accuracy. The model accuracy reflects how closely the kinetic model matches reality. Examples are plotted in Fig.~\ref{fig:model_uncertainty}.

The median results (over 100 simulations) in Table~\ref{tab:model_uncertainty} show that although MPC performs better than the POMDP approach when the model is accurate, its performance lags behind the POMDP approach as the model accuracy decreases. Table~\ref{tab:reward} shows that in the low accuracy scenario, there are even cases where MPC performs worse than a PID controller.

\begin{table}[ht]
\centering
\caption{Median performance of MPC and POMDP approaches relative to PID controller baseline when varying model accuracy (i.e., increasing model uncertainty).}
\begin{tabular}{l c c c c c c}
  \multirow{2}{8em}{Model Accuracy} & \multicolumn{3}{c}{Model Predictive Control} & \multicolumn{3}{c}{POMDP Approach}\\
  \cline{2-7}
  & High & Med & Low & High & Med & Low \\
  \hline
  rel.~recovery $[\Delta\%]$ & -3.6 & -3.8 & -4.3 & -3.1 & -3.3 & -3.1 \\
  rel.~grade $[\Delta\%]$ & +0.4 & +0.5 & +0.9 & +0.4 & +0.7 & +1.9 \\
  rel.~reward $[\Delta \$M/yr]$ & \textbf{+119} & \textbf{+121} & \textbf{+126} & \textbf{+95} & \textbf{+129} & \textbf{+283} \\
\end{tabular}
\label{tab:model_uncertainty}
\end{table}

\subsubsection{Effect of Feedstock Variability}
\label{sec:feedstock_variability}

Now, we consider the effect of feedstock variability on optimization under model uncertainty when the feedstock is fully known (i.e., measured at every timestep). We test different variances of the feedstock composition under a set of grade and recovery surfaces for which, when the feedstock composition is constant, MPC and the POMDP approach have near-equivalent performance. This corresponds roughly to the medium model accuracy case (see Fig.~\ref{fig:model_uncertainty}). (A log variance of -3 can be considered near-constant feedstock, as can be seen in the plots of sample feedstock composition signals in Fig.~\ref{fig:sample_feedstocks}.)

\begin{figure}[htp]
\centering
\includegraphics[trim={0 1.5cm 0 1.5cm},clip,width=0.75\textwidth]{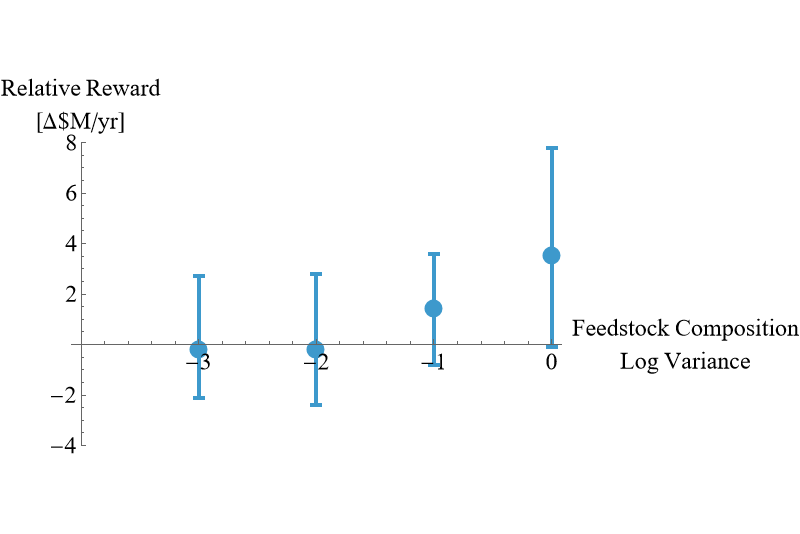}
\caption{Median reward of POMDP approach relative to MPC. Error bars represent 20th and 80th percentiles. Log correlation length is fixed at 2.}
\label{fig:feedstock_variance}
\end{figure}

As can be seen in Fig.~\ref{fig:feedstock_variance}, as the feedstock variance increases, the relative reward of the POMDP approach increases. Detailed results (shown in Table~\ref{tab:feedstock_variability}) indicate that decreasing the feedstock composition correlation length also seems to lead to an increase in the relative reward of the POMDP approach. However, this only occurs at high variance, and the effect is less pronounced.

\subsection{Optimization under State Uncertainty}
\label{sec:state_uncertainty}

Finally, we consider the independent effect of state (i.e., feedstock) uncertainty. We fix the grade and recovery surfaces, and choose a high variance feedstock composition signal. Then, the number of feedstock measurements across the simulation is varied. Taking fewer measurements corresponds to higher state uncertainty. A high, medium, and low model accuracy scenario are each considered.

\begin{figure}[hbtp]
\centering
\begin{subfigure}[c]{0.325\textwidth}
  \includegraphics[width=\textwidth]{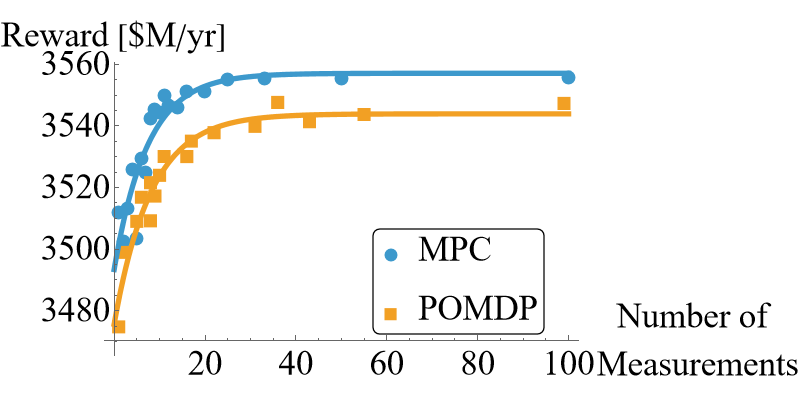}
  \caption{High model accuracy}
  \label{fig:feedstock_measurements_3}
\end{subfigure}
\begin{subfigure}[c]{0.325\textwidth}
  \includegraphics[width=\textwidth]{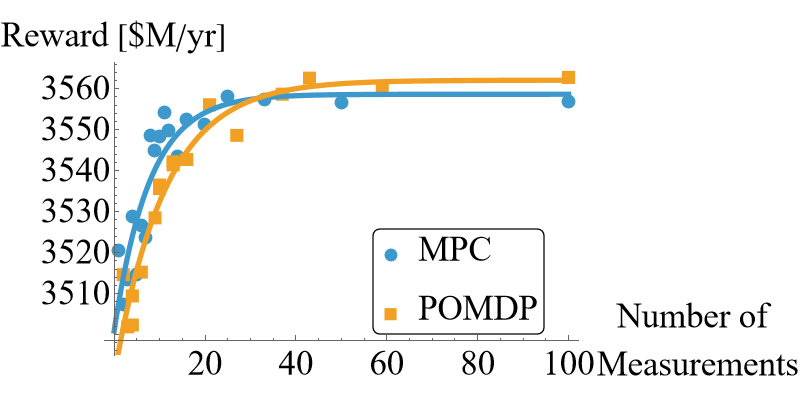}
  \caption{Medium model accuracy}
  \label{fig:feedstock_measurements_2}
\end{subfigure}
\begin{subfigure}[c]{0.325\textwidth}
  \includegraphics[width=\textwidth]{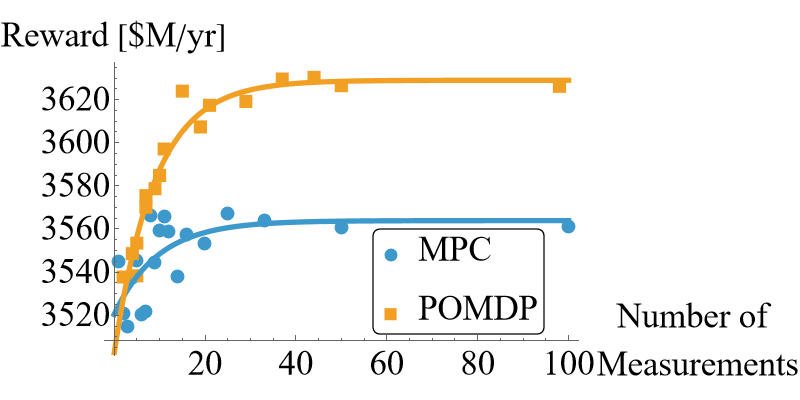}
  \caption{Low model accuracy}
  \label{fig:feedstock_measurements_1}
\end{subfigure} 
\centering
\caption{Performance of MPC and POMDP approaches for a high-variance feedstock composition signal and three fixed grade and recovery surfaces, varying the number of measurements. Lines are exponential curves of best fit.}
\label{fig:feedstock_measurements}
\end{figure}

As shown in Fig.~\ref{fig:feedstock_measurements}, for a high variability feedstock, the POMDP approach improves at a faster rate than MPC as the number of measurements (i.e., information gathered) increases. The POMDP approach is never able to outperform MPC at high model accuracy (consistent with the results in Sec.~\ref{sec:model_uncertainty}), nor in any case with zero measurements. However, after a certain number of measurements in the medium and low accuracy cases, the POMDP approach is able to surpass MPC. For the medium accuracy case, crossover occurs at $n=30$, and for the low accuracy case, crossover occurs at $n=3$ (where $n=$ number of measurements).

\section{Discussion}

\subsection{Analysis of Results}

The test cases presented reveal that an optimization-under-uncertainty approach using a POMDP framework is more successful at handling cases of significant model uncertainty, adjusting to significant feedstock variability under model uncertainty, and utilizing limited information under state (i.e., feedstock) uncertainty.

In Sec.~\ref{sec:model_uncertainty}, we see that for MPC, a high-quality deterministic optimization algorithm whose performance depends heavily on the quality of the model (i.e., having an accurate picture of the system), performance correlates strongly with the accuracy of the model. In the high accuracy case, the predicted optimal region largely overlaps with actual optimal region, so using the kinetic model suffices to inform optimal decision-making. As a result, MPC is not only sufficient, it outperforms the POMDP approach.

But as soon as the model deviates from reality, MPC struggle, while a POMDP approach consistently performs well, even when the flotation model has low accuracy. In the medium accuracy case, there is some overlap in optimal regions, but there is now a new optimal region that is not captured by the kinetic model. In the low accuracy case, there is almost no overlap in optimal regions, so relying on the kinetic model would result in suboptimal decision-making.

(Note that both MPC and the POMDP approach consistently find solutions that result in worse recoveries but better grades than the PID controller. This is because these two approaches are both seeking to optimize the reward, which takes into account operating costs, while the PID controller just seeks to maintain a high grade and recovery setpoint regardless of other factors.)

Even with relatively high model accuracy, small inaccuracies in the model can compound when the feedstock has high variability, as shown in Sec.~\ref{sec:feedstock_variability}. The POMDP approach is able to adjust and better account for feedstock variability by developing a more accurate model over time. These results imply that, in this case, having more accurate information about the process model is more important than having perfect information about the feedstock in optimizing performance---in other words, model uncertainty matters more than feedstock uncertainty.

The importance of model uncertainty over feedstock uncertainty is further supported by the results in Sec.~\ref{sec:state_uncertainty}. The POMDP approach almost always achieves a greater reward than MPC in the low model accuracy case, while the opposite is true in the high model accuracy case, regardless of the level of feedstock uncertainty. If feedstock uncertainty had a greater influence on performance, we would expect to see a crossover of the curves for all depicted model accuracies.

However, feedstock uncertainty is still relevant, which is particularly clear in the medium model accuracy case (see Fig.~\ref{fig:feedstock_measurements_2}). The greater rate of improvement for the POMDP approach as the number of measurements increase indicates that it can translate additional feedstock information into a greater reward increase than MPC can. In other words, a POMDP approach ``learns'' faster than MPC as more data is acquired.

\subsection{Implications for Real Systems}

These results emphasize the power of the POMDP in handling applications with high degrees of uncertainty, especially when the model of the system (i.e., the description of process dynamics) has low accuracy. Note that although these results are for simple test cases, we would expect that in more complex, real systems, there would be significant model uncertainty, which a POMDP approach is better equipped to handle than deterministic optimization.

This approach has immediate relevance for potential application in improving the design-of-experiments of bench-scale flotation cells as well as in optimizing the operation of industrial-scale flotation cells without the need for any retrofitting. And although we use the example of a flotation cell in this work, the framework can be adapted to any process unit, or scaled up for optimization of a flotation circuit, an entire mineral processing circuit, an integrated mine and processing plant, and even an entire vertical mineral supply chain.

Beyond improving existing operations, using a POMDP approach could aid in the design and optimization of versatile, highly-adaptable mineral processing plants that were previously not possible due to feedstock variability and process complexity. Such a processing plant could even obviate the need for blending, as it could adjust operational settings for a wide range of possible feedstocks.

Lastly, the use of a solver based on Monte Carlo tree search ensures that the optimization decisions it makes are interpretable. The goal is to aid mineral processing experts in making decisions, rather than to take over decision-making with AI altogether.

\section{Conclusions}

We have demonstrated that mineral processing can be framed as a problem of optimization-under-uncertainty, presenting a mathematical formulation of a simplified flotation cell using the POMDP framework. A range of synthetic test cases demonstrate the utility of the POMDP approach compared to deterministic approaches like MPC, especially in cases with significant feedstock (i.e., state) and process (i.e., model) uncertainty. Through belief updating, the POMDP formulation is designed to incorporate both feedstock (i.e., state) and process (i.e., model) uncertainty when performing optimization, which enables it to handle conditions of significant uncertainty more readily than deterministic methods such as MPC. Thus, an optimization-under-uncertainty approach is particularly well-suited for optimizing mineral processing.

Our work has presented the following advancements:
\begin{enumerate}
    \item Mineral processing can be framed as a problem of optimization-under-uncertainty, as demonstrated by our mathematical formulation of a simplified flotation cell.
    \item Framing the ultimate goal as optimization, rather than control, is better suited to handling uncertainty in mineral processing. MPC's performance over PID alone emphasizes this.
    \item The representation of an unknown state and model via a belief, and the integration of real-time data collection into process optimization via belief updating, is fundamental to how a POMDP approach models uncertainty and the reduction of uncertainty over time. In other words, an intelligent agent learns a more accurate model of process dynamics and estimation of feedstock variability over time to improve process optimization.
    \item Synthetic test cases confirm that in scenarios with significant feedstock (i.e., state) and process (i.e., model) uncertainty, a POMDP approach performs better than deterministic approaches like MPC.
\end{enumerate}

Future work is needed to apply this approach to real-world test cases. The nearest-term application could be for design-of-experiments of bench-scale flotation. The formulation and code as presented in this paper could be directly applied, along with a few tweaks to add complexity, such as: a more specific, well-developed flotation model, the inclusion of feedstock characteristics beyond an average composition, a larger set of control parameters, and a more case-specific reward function. Similarly, this work could be readily applied to optimizing the operation of an industrial-scale flotation cell, with similar tweaks, as well as swapping out flotation time for feed rate to consider a continuous flotation process. The more fruitful application would be in the design and operation of mineral processing circuits, which could apply the same approach, but would require the development of a new mathematical formulation. We hope that this will inspire future work to improve the efficiency and sustainability of industrial-scale mineral processing facilities.

\section*{Acknowledgments}

This work was funded through the Mineral-X Industrial Affiliates program, which is supported by affiliate members KoBold Metals, Bidra VC, Ero Copper, Fleet Space Technologies, Ideon Technologies, and Xcalibur Smart Mapping. We thank them for their support.

\section*{Conflicts of Interest}

The authors declare that they have no conflict of interest.

\bibliographystyle{elsarticle-num-names} 
\bibliography{references.bib}

@online{wmo2025,
  author = {{World Meteorological Organization (WMO)}},
  title = {{WMO} confirms 2024 as warmest year on record at about 1.55{°C} above pre-industrial level},
  year = 2025,
  url = {https://wmo.int/news/media-centre/wmo-confirms-2024-warmest-year-record-about-155degc-above-pre-industrial-level},
  urldate = {2025-06-06}
}

@article{lee2023synthesis,
  title={Synthesis report of the IPCC Sixth Assessment Report (AR6), Longer report. IPCC.},
  author={Lee, Hoesung and Calvin, Katherine and Dasgupta, Dipak and Krinmer, Gerhard and Mukherji, Aditi and Thorne, Peter and Trisos, Christopher and Romero, Jose and Aldunce, Paulina and Barret, Ko and others},
  year={2023},
  publisher={Intergovernmental Panel on Climate Change (IPCC)}
}

@online{iea2021criticalminerals,
  author       = {{International Energy Agency (IEA)}},
  title        = {The Role of Critical Minerals in Clean Energy Transitions},
  year         = {2021},
  institution  = {International Energy Agency},
  location     = {Paris},
  url          = {https://www.iea.org/reports/the-role-of-critical-minerals-in-clean-energy-transitions},
  note         = {Licence: CC BY 4.0},
  urldate      = {2025-06-06}
}

@incollection{bascur2019processcontrol,
  author       = {Osvaldo Bascur},
  title        = {Process Control and Operational Intelligence},
  booktitle    = {SME Mineral Processing and Extractive Metallurgy Handbook},
  editor       = {Robert C. Dunne and S. Komar Kawatra},
  publisher    = {Society for Mining, Metallurgy, and Exploration (SME)},
  year         = {2019},
  chapter      = {2.7},
  pages        = {277--316},
  location     = {Englewood, CO}
}

@book{bascur2024engineering,
  title={The Engineering Science of Mineral Processing: A Fundamental and Practical Approach},
  author={Bascur, Osvaldo A and others},
  year={2024},
  publisher={CRC Press}
}

@article{jiang2017data,
  title={Data-driven flotation industrial process operational optimal control based on reinforcement learning},
  author={Jiang, Yi and Fan, Jialu and Chai, Tianyou and Li, Jinna and Lewis, Frank L},
  journal={IEEE Transactions on Industrial Informatics},
  volume={14},
  number={5},
  pages={1974--1989},
  year={2017},
  publisher={IEEE}
}

@article{xiang2021recent1,
  title={Recent advances in deep reinforcement learning applications for solving partially observable markov decision processes (pomdp) problems: Part 1—fundamentals and applications in games, robotics and natural language processing},
  author={Xiang, Xuanchen and Foo, Simon},
  journal={Machine Learning and Knowledge Extraction},
  volume={3},
  number={3},
  pages={554--581},
  year={2021},
  publisher={MDPI}
}

@article{xiang2021recent2,
  title={Recent advances in Deep Reinforcement Learning applications for solving Partially Observable Markov Decision Processes (POMDP) problems part 2—applications in transportation, industries, communications and networking and more topics},
  author={Xiang, Xuanchen and Foo, Simon and Zang, Huanyu},
  journal={Machine Learning and Knowledge Extraction},
  volume={3},
  number={4},
  pages={863--878},
  year={2021},
  publisher={MDPI}
}

@article{amini2021design,
  title={Design of cell-based flotation circuits under uncertainty: a techno-economic stochastic optimization},
  author={Amini, Seyed Hassan and Noble, Aaron},
  journal={Minerals},
  volume={11},
  number={5},
  pages={459},
  year={2021},
  publisher={MDPI}
}

@book{amini2017optimization,
  title={Optimization of Mineral Processing Circuit Design under Uncertainty},
  author={Amini, Seyed Hassan},
  year={2017},
  publisher={West Virginia University}
}

@article{koermeroptimization,
  title={Optimization of a Metallurgical Process with Uncertain Dynamics},
  year={2021},
  author={Koermer, Scott and Noble, Aaron}
}

@article{koermer2022bayesian,
  title={Bayesian Methods for Mineral Processing Operations},
  author={Koermer, Scott Carl},
  year={2022},
  publisher={Virginia Tech}
}

@article{arief2025managing,
  title={Managing Geological Uncertainty in Critical Mineral Supply Chains: A POMDP Approach with Application to US Lithium Resources},
  author={Arief, Mansur and Alonso, Yasmine and Oshiro, CJ and Xu, William and Corso, Anthony and Yin, David Zhen and Caers, Jef K and Kochenderfer, Mykel J},
  journal={arXiv preprint arXiv:2502.05690},
  year={2025}
}

@book{kochenderfer2022algorithms,
  title={Algorithms for decision making},
  author={Kochenderfer, Mykel J and Wheeler, Tim A and Wray, Kyle H},
  year={2022},
  publisher={MIT press},
  url={https://algorithmsbook.com/decisionmaking/}
}

@article{silver2010monte,
  title={Monte-Carlo planning in large POMDPs},
  author={Silver, David and Veness, Joel},
  journal={Advances in neural information processing systems},
  volume={23},
  year={2010}
}

@techreport{usgs2024phosphate,
  author       = {{U.S. Geological Survey}},
  title        = {Phosphate Rock},
  institution  = {U.S. Geological Survey},
  series       = {Mineral Commodity Summaries 2024},
  number       = {2024},
  address      = {Reston, VA},
  year         = {2024},
  url          = {https://pubs.usgs.gov/periodicals/mcs2024/mcs2024-phosphate.pdf},
  doi          = {10.3133/mcs2024}
}

@techreport{worldbank2025pink,
  author       = {{World Bank}},
  title        = {World Bank Commodities Price Data (The Pink Sheet): April 2025},
  institution  = {World Bank},
  address      = {Washington, DC},
  year         = {2025},
  month        = {April},
  url          = {https://thedocs.worldbank.org/en/doc/18675f1d1639c7a34d463f59263ba0a2-0050012025/related/CMO-Pink-Sheet-April-2025.pdf}
}

@article{ding2012knowledge,
  title={Knowledge-based global operation of mineral processing under uncertainty},
  author={Ding, Jinliang and Chai, Tianyou and Wang, Hong and Chen, Xinkai},
  journal={IEEE Transactions on Industrial Informatics},
  volume={8},
  number={4},
  pages={849--859},
  year={2012},
  publisher={IEEE}
}

@article{hodouin2001state,
  title={State of the art and challenges in mineral processing control},
  author={Hodouin, Daniel and J{\"a}ms{\"a}-Jounela, S-L and Carvalho, MT and Bergh, Luis},
  journal={Control Engineering Practice},
  volume={9},
  number={9},
  pages={995--1005},
  year={2001},
  publisher={Elsevier}
}

@article{koch2020sequential,
  title={Sequential decision-making in mining and processing based on geometallurgical inputs},
  author={Koch, Pierre-Henri and Rosenkranz, Jan},
  journal={Minerals Engineering},
  volume={149},
  pages={106262},
  year={2020},
  publisher={Elsevier}
}

@article{valikangas2025evaluation,
  title={Evaluation of model uncertainty propagation in mineral process flowsheet designs},
  author={V{\"a}likangas, Henri and Ohenoja, Markku and Brochot, St{\'e}phane and Fern{\'a}ndez, Manuel Gonz{\'a}lez and Ruuska, Jari and Ruusunen, Mika},
  journal={Scandinavian Simulation Society},
  pages={456--463},
  year={2025}
}

@article{mccoy2019machine,
  title={Machine learning applications in minerals processing: A review},
  author={McCoy, John T and Auret, Lidia},
  journal={Minerals Engineering},
  volume={132},
  pages={95--109},
  year={2019},
  publisher={Elsevier}
}

@article{hodouin2011methods,
  title={Methods for automatic control, observation, and optimization in mineral processing plants},
  author={Hodouin, Daniel},
  journal={Journal of Process Control},
  volume={21},
  number={2},
  pages={211--225},
  year={2011},
  publisher={Elsevier}
}

@article{shean2011review,
  title={A review of froth flotation control},
  author={Shean, BJ and Cilliers, JJ},
  journal={International Journal of Mineral Processing},
  volume={100},
  number={3-4},
  pages={57--71},
  year={2011},
  publisher={Elsevier}
}

@article{jovanovic2015contemporary,
  title={Contemporary advanced control techniques for flotation plants with mechanical flotation cells--A review},
  author={Jovanovi{\'c}, Ivana and Miljanovi{\'c}, Igor},
  journal={Minerals Engineering},
  volume={70},
  pages={228--249},
  year={2015},
  publisher={Elsevier}
}

@article{bai2025artificial,
  title={Artificial intelligence of mineral processing process: A review of research progress},
  author={Bai, Zhe and Gao, Peng and Chu, Mansheng and Han, Yuexin and Yuan, Shuai and Tang, Jue and Li, Yanzhao and Shi, Quan and Qiao, Jinghui and He, Jiahao},
  journal={Journal of Environmental Chemical Engineering},
  pages={118313},
  year={2025},
  publisher={Elsevier}
}

\appendix

\newpage
\section{Detailed Results for Low Accuracy Case}
\label{app:model_uncertainty}

\begin{table}[ht]
\centering
\caption{Comparison of reward between different control and optimization approaches.}
\begin{tabular}{l c c c}
  \multirow{2}{4em}{Policy} & \multicolumn{3}{c}{Relative Reward ($\Delta$\$M/yr)}\\
  \cline{2-4}
  & p20 & p50 & p80 \\
  \hline
  PID Controller (Baseline) & - & - & - \\
  Model Predictive Control (MPC) & -14 & \textbf{+126} & +258 \\
  Online Solver (DMU approach) & +188 & \textbf{+283} & +381 \\
\end{tabular}
\label{tab:reward}
\end{table}

\newpage

\section{Sample Feedstock Signals}
\label{app:sample_feedstocks}

\begin{figure}[!h]
\centering
\includegraphics[width=16cm, angle=90]{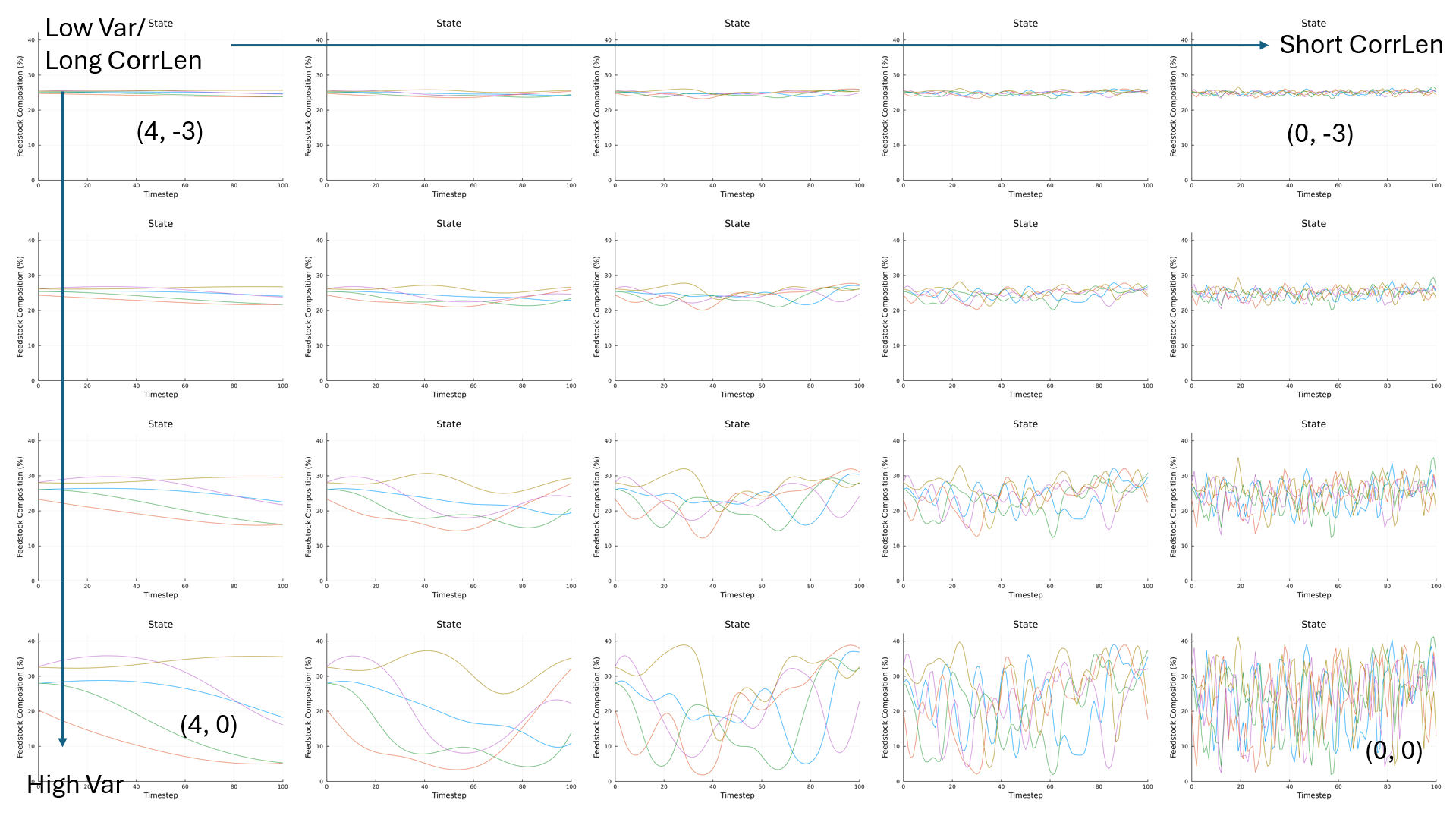}
\caption{Five example feedstock compositions over time with different variances and correlation lengths. The log variance increases from -3 to 0 from left to right, and the log correlation length decreases from 4 to 0 from top to bottom.}
\label{fig:sample_feedstocks}
\end{figure}

\section{Detailed Results for Feedstock Variability}
\label{app:feedstock_variability}

Detailed results for Sec.~\ref{sec:feedstock_variability}.

\begin{table}[htp]
\small
\centering
\caption{Reward of POMDP approach relative to MPC (in $\Delta$\$M/yr) when varying feedstock composition correlation length and variance.}
\begin{tabularx}{\textwidth}{l ccc ccc ccc ccc}
  \multirow{3}{2em}{Log Corr. Len.} & \multicolumn{12}{c}{Feedstock Composition Log Variance}\\
  \cline{2-13}
  & \multicolumn{3}{c}{-3.0} & \multicolumn{3}{c}{-2.0} & \multicolumn{3}{c}{-1.0} & \multicolumn{3}{c}{0.0} \\
  \cline{2-4} \cline{5-7} \cline{8-10} \cline{11-13}
  & p20 & p50 & p80 & p20 & p50 & p80 & p20 & p50 & p80 & p20 & p50 & p80 \\
  \hline
  4.0 & -6.0 & \textbf{0.1} & 5.4 & -5.8 & \textbf{-1.3} & 5.5 & -3.5 & \textbf{0.8} & 6.5 & -2.5 & \textbf{1.6} & 9.8 \\
  3.0 & -3.8 & \textbf{-0.2} & 3.7 & -3.7 & \textbf{0.3} & 2.9 & -2.3 & \textbf{1.2} & 4.5 & -1.7 & \textbf{3.0} & 9.5 \\
  2.0 & -2.1 & \textbf{-0.2} & 2.7 & -2.4 & \textbf{-0.2} & 2.8 & -0.8 & \textbf{1.4} & 3.6 & -0.1 & \textbf{3.5} & 7.8 \\
  1.0 & -2.0 & \textbf{0.1} & 1.8 & -2.5 & \textbf{-0.0} & 1.6 & -0.3 & \textbf{1.5} & 3.9 & 0.9 & \textbf{4.4} & 6.6 \\
  0.0 & -1.4 & \textbf{0.2} & 1.9 & -1.5 & \textbf{0.1} & 1.2 & 0.0 & \textbf{1.7} & 3.1 & 1.3 & \textbf{3.8} & 5.6
\end{tabularx}
\label{tab:feedstock_variability}
\end{table}

\newpage
\section{Action Space Granularity}
\label{app:action_space}

The size and granularity of the action space (i.e., control parameters) affect the results, since a larger action space cannot be explored as efficiently by a Monte Carlo tree search algorithm. In this paper, we use an action grid spacing of $[0.5, 5.0]$ (i.e, consider flotation times with step size 0.5 minutes and air flow rates with step size 5.0 L/hr) for all results, since it best reflects the most fine-grained control settings that are still realistic. Additional testing in Table~\ref{tab:action_space} shows that for larger action spaces (i.e., smaller step sizes), more model uncertainty is necessary for the POMDP approach to perform better than MPC.

\begin{table}[ht]
\centering
\caption{Reward of POMDP approach relative to MPC (in $\Delta$\$M/yr) when varying action space granularity at different levels of grade and recovery error variance.}
\begin{tabularx}{0.8\textwidth}{l *{4}{Y}}
  \multirow{2}{6em}{Action Space Step Size} & \multicolumn{4}{c}{Grade and Recovery Log Variance}\\
  \cline{2-5}
  & -3.0 & -2.0 & -1.0 & 0.0 \\
  \hline
  $[0.1, 1.0]$ & & & -9 & +94 \\
  $[0.25, 2.5]$ & -14 & -4 & +21 & +127 \\
  $[0.5, 5.0]$ & -5 & +9 & +78 & \\
\end{tabularx}
\label{tab:action_space}
\end{table}

\end{document}